\pdfoutput=1
\documentclass[letterpaper]{article} 
\usepackage{aaai2026}  
\usepackage{times}  
\usepackage{helvet}  
\usepackage{courier}  
\usepackage[hyphens]{url}  
\usepackage{graphicx} 
\def\UrlFont{\rm}  
\usepackage{natbib}  
\usepackage{caption} 
\usepackage{url}            
\usepackage{booktabs}       
\usepackage{nicefrac}       
\usepackage{microtype}      
\usepackage{tabularx}       
\usepackage{booktabs}       
\usepackage{longtable}      
\usepackage{subcaption}     

\newcommand{\boldurl}[1]{{\def\UrlFont{\rmfamily\bfseries}\url{#1}}}

\title{``Death by a thousand taxonomies?'': AI Risk Classification In Practice} 
\author{
    Glen Berman\textsuperscript{\rm 1}, Ned Cooper\textsuperscript{\rm 2}, Angel Hsing-Chi Hwang\textsuperscript{\rm 3}, Wesley Hanwen Deng\textsuperscript{\rm 4}, Renee Shelby\textsuperscript{\rm 5}, Ben Hutchinson\textsuperscript{\rm 5}
}
\affiliations{
    \textsuperscript{\rm 1}Australian National University\\
    \textsuperscript{\rm 2}Cornell University\\
    \textsuperscript{\rm 3}University of Southern California\\
    \textsuperscript{\rm 4}Microsoft Research,\\
    \textsuperscript{\rm 5}Google Research\\
    glen.berman@anu.edu.au, ned.cooper@cornell.edu, angel.hwang@usc.edu, wesleydeng@microsoft.com, reneeshelby@google.com, benhutch@google.com
}

\begin{document}

\maketitle

\begin{abstract}
    The harms in which AI is implicated range in nature and scope from unsafe user interactions through to the societal-wide consequences of AI adoption. Classification of the diverse risks of AI is foundational to AI governance: regulators, technology firms, and policymakers need structured accounts of risk upon which to act. Researchers and practitioners have accordingly developed many Sociotechnical Outcome Taxonomies (SOT). This paper presents an empirical study of SOT development and use, drawing on 25 interviews with researchers and practitioners across industry, academia, civil society, and government. We find SOT are weakly integrated into AI governance processes, and identify two features of SOT design and use that explain why. First, the design choices through which SOT produce structured representations of the complex problem space of AI risks tend to be invisible to downstream taxonomy users. Those users treat the resulting categories as exhaustive accounts of risk rather than as interpretive aids. Second, SOT typically enumerate harms without linking them to decision points or actors implicated in their occurrence, leaving accountability difficult to assign. We close with design recommendations for SOT developers and users, and argue realising the potential of SOT requires governance infrastructure that does not yet exist.
\end{abstract}

\section{Introduction}

Researchers, policymakers, and industry practitioners have 
developed structured classifications of the risks, harms, and impacts associated with Artificial Intelligence (AI) systems \citep[e.g.,][]{weidinger2022TaxonomyRisksPosed, hutiri2024NotMyVoice, dominguezhernandez2024MappingIndividualSocial, chan2023HarmsIncreasinglyAgentica}. These artefacts --- variously termed taxonomies, typologies, repositories, or frameworks --- are designed to serve as boundary objects, rendering the social consequences of AI development legible, enumerable, and governable \citep{star2010ThisNotBoundary}. We refer to these artefacts as \textit{sociotechnical outcome taxonomies} (SOT). SOT are intended to be foundational to AI governance, referring to the practices, methods, policies, and regulations used to manage the development, deployment, and monitoring of AI technologies \citep{batool2025ai}. SOT form part of an emerging set of artefacts responding to civil society demands for accountability in AI development. Specifically, SOT translate the concerns of stakeholders in AI development, internal and external to AI firms, into actionable schemas for product teams.


SOT have proliferated over the past decade. Early efforts, primarily from academia, 
established conceptual foundations, such as distinguishing allocational from representational harms in machine learning \citep{barocas2017ProblemBiasAllocative}, identifying stakeholder-centred risks in language technologies \citep{bender2019TypologyEthicalRisks}, and highlighting distinctions between process- and outcomes-based approaches to harms \citep[e.g.,][]{binns2018FairnessMachineLearning}. As foundation models and generative AI systems scaled commercially, industry research labs became significant SOT producers, developing taxonomies oriented towards deployed systems \citep{weidinger2022TaxonomyRisksPosed, bird2023TypologyRisksGenerative, shelby2023SociotechnicalHarmsAlgorithmic, solaiman2023EvaluatingSocialImpacta}. Parallel efforts emerged from civil society, intergovernmental bodies, and national standards institutions, drawing on rights-based framings, regulatory alignment, and community-centred risk identification \citep{TaxonomyHumanRights, calvino2024SectoralTaxonomyAi, vassilev2024AdversarialMachineLearning, abercrombie2024CollaborativeHumanCentredTaxonomy}. More recently, academic and civil society projects have focused on mapping the landscape \citep{slattery2026AIRiskRepository, arnold2024IntroducingAIGovernance}. Taxonomising has thus become central to Responsible AI (RAI) practice.


Little empirical research, however, has examined how SOT are developed, interpreted, and used in practice, or how their impact might be evaluated. Currently, we lack a systematic understanding of how SOT developers conceptualise their work, how practitioners select and operationalise SOT within AI development, and how these artefacts mediate broader governance dynamics. Such understanding is necessary to inform efforts to standardise SOT or strengthen their role in AI governance. Examining their use in practice is important because classification is never neutral: categorisation systems stabilise particular problem framings, rendering some phenomena visible and actionable while marginalising others \citep{bowker2000SortingThingsOut}. 
Accordingly, this paper critically reflects on SOT development, proliferation, and use to ask:
\begin{description}
    \item [\textbf{RQ1}] How are SOT developed and put to use within the AI product development and deployment pipeline?
    \item [\textbf{RQ2}] What is the role of SOT in the broader project of AI governance, and how might this role be strengthened?
\end{description}

To answer these questions, we conducted a qualitative interview study of 25 researchers and practitioners across industry, academia, and civil society who have developed or used SOT. Through reflexive thematic analysis, we examine how SOT are imagined, constructed, operationalised, and contested.
Our analysis is informed by Science and Technology Studies (STS) scholarship on classification and standardisation \citep[e.g.,][]{bowker2000SortingThingsOut, star1989StructureIllStructuredSolutions, timmermans2010WorldStandardsNot}. 
We conceptualise SOT as \textit{proto-standards}, that is, standardisation efforts that seek to stabilise the problem space of AI harms. Through what \citet{bowker2000SortingThingsOut} describe as ``infrastructural inversion,'' we shift analytical attention from the harms enumerated within taxonomies to the classificatory infrastructure itself. This perspective foregrounds the political work of categorisation, specifically the negotiation of commensurability, creation of residual categories, and temporal challenges of maintaining standards in rapidly evolving technological environments. We complement this framing with insights from Human-Computer Interaction (HCI) research on the frictions of operationalising RAI.
In providing one of the first qualitative studies of the development and use of SOT, our work makes two contributions. First, we articulate design and use considerations for SOT developers and AI practitioners. 
Second, we describe how SOT can be better embedded within the broader AI governance landscape. 


\section{Related work}


\subsection{Sociotechnical outcomes taxonomies}

AI risks, harms and outcomes are products of complex interactions between sociotechnical systems and multilevel social phenomena, ranging from individual user interactions to the societal distributions of resources and power. Accordingly, they resist straightforward description and categorisation. 

The development of SOT emerged as a pragmatic response to the operationalisation gap in AI governance \citep{mittelstadt2019AIEthicsToo, morley2020WhatHowInitial}. Technology firms, in particular, promulgated high-level ethical principles signalling to internal teams and external stakeholders how they would research, develop, and deploy AI technologies, while mitigating risks and avoiding social harms \citep{jobin2019GlobalLandscapeAI, greene2019BetterNicerClearer}. Translating these abstract principles into engineering constraints, however, has proven challenging. It requires practitioners to develop formal definitions of high-level concepts, such as \textit{fairness} \citep{narayanan2018TranslationTutorial21}, to guide their operationalisation \citep{mittelstadt2019AIEthicsToo, selbst2019FairnessAbstractionSociotechnical, jacobs2021MeasurementFairness}. 
Researchers and practitioners have introduced SOT as translation artefacts \citep{star1989InstitutionalEcologyTranslations}, to both map the terrain of sociotechnical hazards associated with AI technologies and to stabilise this terrain in a way that is navigable for, and reconcilable with, AI product development. Much like checklists, SOT can ``provide organizational infrastructure for formalizing ad-hoc processes and empowering individual advocates'' \citep[p.10]{madaio2020CoDesigningChecklistsUnderstanda}. In this sense, SOT are intermediate governance artefacts that translate broad normative commitments into categories that can be discussed, documented, prioritised, and measured.

As AI technologies have expanded from research-oriented language models into consumer-facing, multimodal, and multi-model generative AI systems, SOT have proliferated and diversified. As of May 2026, there are over 70 SOT \cite{slattery2026AIRiskRepository}; Table \ref{tab:published_sots} in the Appendix provides a non-exhaustive list of published SOT to demonstrate this diversity. Initial efforts focused on the societal impacts of Natural Language Processing (NLP) \citep{hovy2016SocialImpactNatural, weidinger2022TaxonomyRisksPosed}. Subsequent work extended this classificatory effort across system types and analytical dimensions. For example, \citet{slaughter2020AlgorithmsEconomicJustice} catalogued consumer-facing harms from algorithmic decision-making, while \citet{shelby2023SociotechnicalHarmsAlgorithmic} synthesised harms across algorithmic systems more generally, distinguishing between types of harm and the levels at which they occur. Other research refined boundaries between present harms and anticipated risks \citep{weidinger2022TaxonomyRisksPosed}, 
distinguishing technical evaluations of base AI systems from evaluations of impacts on people and society \citep{solaiman2023EvaluatingSocialImpacta}, and mapped individual, social, and biospheric levels of analysis \citep{dominguezhernandez2024MappingIndividualSocial}. 

SOT efforts have also adapted to new AI modalities, including text-to-image \citep{bird2023TypologyRisksGenerative} and speech generation \citep{hutiri2024NotMyVoice}. These efforts now address specific domains and populations --- such as clinical and public health \citep{golpayegani2022TaxonomyAIRisks, Zhou-public-health-taxonomy-2025}, disability contexts \citep{wang2025TaxonomyAlgorithmicHarms}, and youth \citep{Yu-youth-risk-2025, khoo-minor-bench-2025} --- as well as emerging categories of harm, including psychological, social, and relational harms \citep{Alberts-2024-cconversational, Alabed_AI_relationship_2024, knox-harmful-trait-ai-companions-2025, Zhang-dark-side-2025} and privacy and surveillance risks \citep{Lee-privacy-risks-2024, Gumusel-privacy-harm-2024}. Many emerging-harm taxonomies adopt more inductive methodologies than earlier work, grounding harm categories in lived experiences, stakeholder narratives, and user-reported impacts \citep{Chandra-lived-experience-2025, Archiwaranguprok-case-informed-harm-2025, zhu-user-discourse-2026, li-user-reported-risk-2025}. This methodological shift is occurring in contexts where AI capabilities and social practices are co-evolving rapidly and where harms are not yet well represented in academic literature.

\subsection{Taxonomies as AI governance infrastructure}

SOT are governance instruments that coordinate work across distributed teams, organisations, and regulatory domains. Drawing on prior work, we identify three interrelated functions SOT may perform within AI governance: (1) structuring internal risk deliberation, (2) embedding harm categories into documentation and audit processes, and (3) mediating translation between organisational practices and regulatory regimes. Our empirical findings extend and complicate the brief description of these functions below.

\subsubsection{Structuring risk deliberation}

SOT support internal risk deliberation by providing granular structuring of RAI concerns. High-level principles demarcate areas of ethical salience for AI product teams, but leave open critical questions of interpretation, prioritisation, and implementation \citep{mittelstadt2019AIEthicsToo}. 
SOT concretise these areas by disaggregating high-level concerns into enumerated categories of impact, risk, or harm. In doing so, they create reference points that researchers and practitioners can use during design reviews, evaluation planning, impact assessments, and AI audits. Empirical studies of industry practice indicate that practitioners often lack shared language for discussing sociotechnical harms across product, policy, and legal teams, and face structural barriers when integrating fairness and responsibility concerns into existing workflows \citep{holstein2019ImprovingFairnessMachine, rakova2021WhereResponsibleAI, madaio2022AssessingFairnessAI}. Toolkits and internal governance artefacts can shape how such concerns are framed and by whom they are raised \citep{wongSeeingToolkitHow2023}. \citet{madaio2020CoDesigningChecklistsUnderstanda,madaio2022AssessingFairnessAI}, for example, show that structured artefacts, such as fairness checklists or review templates, facilitate cross-functional communication by making implicit assumptions explicit and routinising some forms of ethical inquiry. SOT operate in a similar manner: they specify what kinds of impacts ought to be considered, and potentially how different impacts may be mitigated. In doing so, SOT implicitly define which impacts require measurement, and they provide categories through which concerns can be documented and communicated to other stakeholders. Ultimately, SOT shape how harms become administratively visible within AI governance, echoing broader concerns about problem framing in sociotechnical systems \citep{selbst2019FairnessAbstractionSociotechnical}.

\subsubsection{Interfacing with AI governance}

SOT acquire infrastructural significance through the documentation and audit processes into which they are embedded.
Transparency documentation frameworks such as model cards \citep{mitchellModelCard2022} and datasheets \citep{gebruDatasheetsDatasets2018} formalise expectations about which ethical considerations technologists should record and report. SOT respond to, and mediate, these expectations by structuring how such considerations are surfaced within companies and how external stakeholders can challenge inadequate consideration. Internal audit frameworks similarly rely on structured risk classification. \citeauthor{rajiClosingAIAccountability2020}'s \citeyearpar{rajiClosingAIAccountability2020} lifecycle auditing model envisions harm identification and mitigation as a staged review process producing an `ethical risk analysis chart,' and Algorithmic Impact Assessments (AIAs) extend this logic by formalising anticipatory risk evaluation \citep{metcalf2021AlgorithmicImpactAssessments, selbst2021InstitutionalViewAlgorithmic}. SOT interface with audits and AIAs, working together to conceptually (through the SOT) and administratively (through the audit or AIAs) define the space of ethically salient concerns and assign responsibility for their management. Yet, documentation alone does not ensure substantive action.
\citet{costanza-chock2022WhoAuditsAuditors} show that shared standards for audit quality and comparability are underdeveloped, and most evaluation criteria for RAI tool development focus on usability rather than demonstrable harm reduction \citep{bermanScopingStudyEvaluation2024}. Consequently, documentation and governance processes may enable the performance of transparency and responsibility without actually requiring it \citep{greene2019BetterNicerClearer}. 

\subsubsection{Moving across internal and external boundaries}

SOT also operate at the interface between organisational practice and regulatory regimes. Risk-based regulatory frameworks, such as the \textit{EU AI Act}, depend on categorical distinctions that organisations must interpret and implement through conformity assessment, technical documentation, and reliance on harmonised standards \citep{veale2021DemystifyingDraftEU}. Data protection laws, on the other hand, compel organisations to articulate risks in structured formats \citep{kaminski2021AlgorithmicImpactAssessments, metcalf2021AlgorithmicImpactAssessments}.
Within this broader standardisation environment, SOT operate as intermediary devices, aligning legal risk categories with internal evaluation metrics and governance procedures, consistent with broader accounts of how standards mediate between regulatory norms and organisational practice \citep{timmermans2010WorldStandardsNot}. AI governance increasingly proceeds through documentation, standardisation, and categorical alignment, with SOT mediating among these.



\subsection{Taxonomy proliferation and challenges}

While the proliferation of SOT reflects growing attention to AI-related harms, their divergences introduce instability in AI governance processes. Existing scholarship identifies coordination challenges, participatory burdens, and missing guidance for SOT use. These challenges produce a complex and unstable landscape of SOT adoption. Yet, limited empirical research examines how SOT are operationalised within AI governance processes. Our interview study addresses this gap by examining how SOT are developed, operationalised, and contested \textit{in situ}.

\subsubsection{Fragmentation and selection ambiguity}

Recent large-scale reviews of sociotechnical harms \citep{shelby2023SociotechnicalHarmsAlgorithmic} and meta-repository initiatives \citep{slattery2026AIRiskRepository} demonstrate overlap and divergence in the scope of SOT, the terminology used within, and the level of abstraction around which they are organised. SOT users must choose between partially overlapping frameworks without established criteria for determining which is most appropriate for a given system, sector, or organisational context. Fragmentation may also enable strategic evasion of AI governance, wherein organisations select SOT that are narrower in scope or omit particular risks --- akin to ``ethics washing'' in other AI governance practices \citep{bietti2020EthicsWashingEthics, arnold2024IntroducingAIGovernance}.

\subsubsection{Participatory burdens}

Many SOT emphasise participatory development processes \citep[e.g.,][]{abercrombie2024CollaborativeHumanCentredTaxonomy}. Such approaches are often normatively justified as enhancing legitimacy and inclusivity. However, in a context where SOT are proliferating, they may create cumulative burdens for community advocates and stakeholders asked to participate in their development. Critical scholarship on design justice and participatory governance highlights the risks of extractive engagement, consultation fatigue, and the uneven distribution of labour onto communities and individuals with limited institutional power \citep{birhane2022PowerPeopleOpportunities, cooper2022SystematicReviewThematic, costanza-chock2020DesignJusticeCommunityled, rakova2021WhereResponsibleAI}. 
When multiple SOT are developed in parallel, potentially consulting overlapping communities, there may be little coordination to mitigate repeated engagement demands or to share insights across initiatives.

\subsubsection{Guidance for use}

While SOT provide structured enumerations of harms or risks, they rarely specify how they should be selected, combined, or operationalised in practice. Comprehensive taxonomies \citep[e.g.,][]{weidinger2022TaxonomyRisksPosed, slattery2026AIRiskRepository} describe categories and subcategories but typically do not offer decision frameworks for prioritising harms or resolving trade-offs between competing impacts. 
Decisions about which harms to prioritise, how to weigh conflicting impacts, and how to determine when an SOT has been sufficiently applied remain largely under-specified.

\section{Method}

Through expert interviews, we explored practices and perspectives about taxonomising societal impacts of AI. 

\subsection{Recruitment}

We recruited participants with experience developing or using SOT. Contributors to SOT were identified through the literature summarised in the preceding section and the professional networks of the research team. 
Two sets of inclusion criteria were used:
\begin{itemize}
    \item \textbf{Taxonomy developers}: Contributors to the development of an SOT designed to enable systematic thinking about the potential consequences of deploying an AI system or AI technology. We interpreted both ``contribution'' (e.g., co-authorship, advisory roles) and ``AI'' broadly: if an SOT used the term AI or associated terms (e.g., GenAI, Frontier AI), we considered it relevant.
    \item \textbf{Taxonomy users}: Individuals who have engaged with SOT in a work or research setting. Engagement could take many forms, from using an SOT in an evaluation of an AI system to using an SOT to inform research. Participants needed some prior familiarity with an SOT (e.g., its layout, scope, and purpose) but did not need to be deeply familiar with multiple SOT.
\end{itemize}
Through email outreach, we recruited 25 participants from industry, academia, and civil society or government, across five regions (see the summary in Table \ref{tab:participant_summary} and further participant details in the Appendix). Among the participants, 17 had contributed to the development of one or more SOT (representing over 20 SOT collectively), and eight had engaged with taxonomies primarily as users in RAI evaluation contexts.

\begin{table}[h]
\centering
\setlength{\tabcolsep}{3pt}
\begin{tabular}{>{\raggedright\arraybackslash}p{2.2cm}ccccc}
\toprule
Sector & Europe & UK & N. America & Other & Total \\
\midrule
Academia            & 0 & 2 & 6 & 0 & 8 \\
Industry            & 2 & 0 & 8 & 0 & 10 \\
Civil Soc./Gov.  & 1 & 1 & 4 & 1 & 7 \\
\midrule
Total               & 3 & 3 & 18 & 1 & 25 \\
\bottomrule
\end{tabular}
\caption{Participants by sector \& geography ($n=25$).}
\label{tab:participant_summary}
\end{table}

\subsection{Data Collection}

We developed two versions of a semi-structured interview protocol, one for SOT developers and one for SOT users, and structured both protocols around two themes:
\begin{enumerate}
    \item \textbf{The development or use of specific SOT}: For developers, this focused on questions about purpose, envisaged use cases, target audiences, and design decisions. For users, this included questions about how they selected SOT for particular tasks, what challenges they encountered when using it, and what made an SOT useful or not.
    \item \textbf{The broader role of SOT in tracking and managing societal outcomes of AI deployments}: This included how SOT fit within the RAI ecosystem and how societal impact is understood and measured in participants' organisations.
\end{enumerate}
All interviews were conducted online, lasted approximately 60 minutes, and were recorded with consent, transcribed, and de-identified. Participants were compensated with gift cards and offered the opportunity to review direct quotes before publication. The project was approved by the ANU IRB.

\subsection{Data analysis}

We analysed the transcripts using reflexive thematic analysis (RTA) \citep{Braun2006Using, Braun2021Size}. This approach foregrounds the influence of researchers in interpreting data, encouraging them to reflect on that influence as they develop and refine codes, and it aims for rich interpretations rather than a consensus of meaning \cite{Byrne2022Worked}. 
All authors participated in the analysis, meeting regularly as a team over six months to review transcripts and inductively develop low-level codes based on the semantic content of each interview. 
Discussions at each meeting compared interpretations, surfaced disagreements, and challenged assumptions that our shared familiarity with the domain could lead us to overlook. 

We grouped low-level codes into higher-level codes on a digital whiteboard; one author led the initial grouping before the broader team progressively developed these into themes through multiple rounds of discussion and refinement. See the Appendix for the codebook produced through this analysis. The distinction between SOT developers and users that we envisaged in our interview design proved difficult to maintain during analysis: several participants recruited as developers described drawing on existing SOT in their own work, while those recruited as users were often simultaneously developing internal SOT. 
We therefore report differences between industry-based participants and those in other sectors, but not between nominal `developers' and `users.'

\subsection{Positionality Statement}

Our team is interdisciplinary and inter-sectoral. Three authors hold industry affiliations and three are based in academia, with disciplinary backgrounds spanning NLP, HCI, and STS. Four authors are based in the United States, two in Australia. All authors are active in the RAI research community. One author led the development of an SOT; others have contributed to the development of SOT or used them in evaluation work. We recognise SOT as one of the few RAI artefacts with uptake across research, policy, and industry, while also being aware of widespread barriers to effective RAI adoption in practice. Our analysis proceeds from an intermediary position, neither disinterested outsiders to questions of SOT development nor entirely insiders to SOT adoption within technology firms.

\section{Findings}

Participants positioned SOT as foundational to the broader project of RAI. In their accounts, SOT enable product teams, policy makers, regulators, activists, and community groups to map and navigate the complex problem space of AI risks and impacts. SOT are ``critical scaffolding'' [C10]\footnote{We refer to participants using A (academia), I (industry) or C (civil society or government) and an ID number (e.g. A1). Several participants have moved between sectors during their career; we display only their  affiliation at the time of this study.} for developing a shared vocabulary to inform risk analysis, impact assessment, and harm mitigation efforts, fostering accountability for AI deployment decisions. While participants' evidence of SOT adoption was largely anecdotal, their reports of interest in SOT [A11, C12, I7, I8, I20] indicate the potential. Participants were cognisant of the challenges of SOT proliferation, adoption, and integration into AI governance, but the opportunity for SOT to be foundational to the development of robust AI governance regimes was nonetheless clear to them [A9, A11, C10, C19, I2]. Yet, participant descriptions of how SOT are used in practice demonstrate our first finding: \textbf{SOT are not well integrated into AI governance processes}, with little evidence of SOT being used to inform AI deployment decisions or the monitoring of AI impacts.

Participants' responses highlighted two significant barriers to realising the potential of SOT, constituting our second and third findings. To structure the landscape of AI risks, \textbf{SOT inherently reduce the complexity of the problem space}.
Enacting this reduction requires resolving design tensions --- most significantly the ``altitude problem'' [I21], wherein SOT developers must choose the right level of abstraction at which to define AI risks or impacts. 
Participants observed that AI risks and impacts are not natural phenomena to be mapped, but rather sociotechnical outcomes whose causes must be explicated to scaffold substantive and targeted AI governance interventions. Yet, \textbf{incorporating causal analysis into the development of SOT has proved challenging}, and most existing SOT do not attempt to identify the decision points or critical actors who may be responsible for the catalogued risks occurring.

Participants advocated for a range of approaches to addressing these barriers. Our final finding reports \textbf{three recommendations for improving SOT development and adoption}, on which participants converged: linking harm categories in SOT to decision-points in product development, strengthening external requirements on technology firms to ensure monitoring of risks enumerated in SOT, and improving coordination across currently overlapping and sometimes divergent SOT frameworks. Together, these recommendations signal a move towards standardisation across SOT, although they raise the question of where within the AI ecosystem, from a normative perspective, SOT development should be led. We return to this question in our Discussion.

\subsection{Gaps in integrating SOT into AI governance}

Across participants' accounts, the integration of SOT into AI governance remains partial and uneven. This is visible in two strands of our data, which we examine in turn: the proliferation of SOT in the absence of coordination across them, and the limited visibility participants reported into the adoption of SOT and the outcomes SOT attempt to anticipate.

\subsubsection{Proliferation without coordination}

Proliferation is a condition practitioners must navigate rather than a resource they can draw on. The multiplicity of SOT creates ``analysis paralysis'' [I8], with attention spent grappling with subtly different and often incompatible classifications rather than acting on the harms those classifications enumerate [I4, A15, I16]. I14 reported that ``there are so many of them'' that ``choosing your approach'' has itself become ``a limitation of taxonomies.'' Organisationally, proliferation enables strategic selection of SOT while confounding efforts to standardise harm classification. As A11 observed, ``the ways that different AI people or AI organisations or companies frame [risks] are sometimes suspiciously narrow.'' I16 located the consequence in the absence of coordinated alternatives: fragmentation drives ``every company\ldots\ to create their own'' SOT. Responses were adaptive rather than corrective and, in aggregate, exacerbated proliferation. Rather than waiting for coordination, participants built local workarounds, ``reformulating what has been said'' in existing SOT [A15]. I4 described creating ``mappings between different taxonomies, so at least all of the work you've done under taxonomy A will still be relevant.'' These workarounds enabled participants to situate their work within a fragmented landscape, but did so by adding further layers to it. I8 described the resulting condition: ``you just need to act and you don't know how or what to act upon.''

\subsubsection{Adoption and outcomes without visibility}

Beyond the navigational challenges of proliferation, SOT developers have little visibility into how their SOT are used. The absence of feedback channels surfaces directly: ``I unfortunately have no evidence whether people have found it useful'' [I20]; there is little ``insight into how [developers] actually do, or if they actually do,'' use them [I14]. I14 and I20 are both based in industry, so their lack of visibility into SOT use is particularly striking, though it should not be misread as a lack of interest in SOT utility. Where evidence of use did exist, it was anecdotal. I8 described an SOT they contributed to as ``quite widely used\ldots\ I mean, not all the time.'' Without infrastructure that would track adoption systematically, neither the developers of SOT nor those evaluating their utility can diagnose where integration is succeeding and where it is failing. The same absence operates downstream, at the level of the harms SOT are intended to anticipate. There is little evidence that risks enumerated in SOT are systematically monitored after deployment. An aspirational adverse event reporting pipeline is ``like a fantasy'' [I8], while C17 reported that post-launch monitoring ``consistently gets de-prioritised.'' Together, these visibility gaps mean that the foundational role participants ascribed to SOT in the broader project of AI governance cannot be empirically assessed from within the current ecosystem.

\subsection{Seeking clarity while navigating reductionism}

Participants described SOT as responding to the complexity of AI impacts by imposing a structure on the problem space [C17, I8, I25, I4, A5, I2, A9, A11, C19, A24]. I8, for example, observed that ``you have a bunch of phenomena that you're interested and confused by\ldots\ you're trying to impose some kind of structure onto this.'' By imposing structure, participants framed SOT as enabling understanding of the problem space and foreshadowing action.

This frame of ``imposing structure'' reflects an awareness of the subjectivity inherent in developing classificatory schemas. This subjectivity begins
in the composition of the SOT development team itself, which may ``naturally narrow the scope'' [A1]. As I25 opined: ``If the taxonomy is created primarily from the\ldots\ circle jerk of computer science, then it's not necessarily actually mapping the full breadth of risk.''
Counterexamples to I25's account exist: SOT development teams spanning disciplinary or institutional boundaries formed shared vocabularies and deepened their appreciation of AI risks through the development process [C10, I2, I7, I8]. However, as we explore below, this awareness is in tension with the desire for SOT that AI practitioners can easily interpret and act on. That desire can incentivise reductionist design choices that simplify the problem space in ways that serve some audiences better than others.

\subsubsection{Reduction treated as description}

The structuring work SOT do is necessarily reductive: a navigable category schema cannot, by definition, preserve all the specificity of the phenomena it organises.
Referring to deepfake harms, I20 illustrated how individual categories collapse phenomena that remain heterogeneous in practice:
\begin{quote}
    The kind of harm that happens if a deepfake is used to fool a grandmother [is different] to the kind of harm that happens with the Marco Rubio thing that\ldots\ happened on a political level\ldots\ to the kind of harm that happens if a CEO's voice is taken. They're actually quite different.
\end{quote}
The single category of ``deepfake harm'' offers structure for those acting on the SOT, but only by collapsing distinctions that matter for who is responsible, who is harmed, and what mitigation might involve. The category risks communicating more clarity than the underlying phenomena warrant.

The simplification of AI risks has unintended consequences for the way AI governance processes recognise and address harms. Several participants recognised the sociopolitical consequences of the choices this reduction entails:
\begin{quote}
    At the end of the day, taxonomies are language creation\ldots\ if language and categorisation is power, taxonomies hold power. Because if you create [a] taxonomy and you leave out categories intentionally, those categories won't get measured. [I20]
\end{quote}
I8 named the corollary risk: ``the danger, of course, is always that you start to mistake your abstraction for reality\ldots\ people think the taxonomy is real.'' In both accounts, the reductive work of an SOT is interpretive, not descriptive. The risk is not that SOT simplify per se, but that the simplifications come to be treated as descriptions of the world rather than as one possible structuring of it. 
 
\subsubsection{Usability concerns guide navigation of design tensions}

SOT enable discussion and debate about AI risks by establishing a shared vocabulary. The framework of an SOT, however, must overcome what I21 termed the ``altitude problem:''
\begin{quote}
    Having a ridiculously high branching factor [or] a very flat taxonomy is not that helpful. You need to have some degree of nice categorical structuring to it. But how do you decide which elements can be promoted to what level within the taxonomy?
\end{quote}
The risk here is the failure to meaningfully clarify the problem space; neither exhaustive lists nor high-level categories enable interpretation of AI risks. C17 
similarly noted that the same categories mean different things to different actors:
\begin{quote}
    The same words mean different things [to all the stakeholders], and you need to be able to root the [SOT] in concrete examples.
\end{quote}
The criterion for determining the right `altitude,' across accounts, is thus ease of interpretation for those expected to act on the SOT [A6, A3, A15, C17].

Some risks are likely easier to describe with ``concrete examples'' than others, which may lead to imbalances across SOT categories. Optimising for interpretability therefore has two risks for SOT: inviting reductionist interpretations of AI risks and impacts, and losing fidelity to the heterogeneity of experiences within a category or across affected communities.

The altitude problem is further complicated by the need for SOT to retain their relevance across successive generations of AI technologies. As A9 observed, a ``real difficulty with taxonomies of technology is that\ldots\ if they're too deep, they become redundant really quickly.'' Thinking of SOT as `living' documents --- which evolve over time while retaining their core structure --- resolved tensions between flexibility and stability for several participants [I4, I16, I20, I21].
I16 described how precise risk definitions may be less useful than a balance of flexibility and stability that allows product teams to intuit risks and move quickly to action. The pace of SOT change is thus calibrated to practitioner workflow, 
without consideration of the needs of other stakeholders.


\subsection{Mitigating risks requires addressing causality}

The purpose of SOT, for many participants, is interpretive rather than declarative [C10, I8, I7]. A15 termed this ``the procedural value of taxonomising,'' which ``maybe exists irrespective of whether or not at the end of it you try to publish a paper.'' I2 framed the SOT they contributed to as ``a sort of reflection, an initial tool embedded in the process before we go to a more formal process.'' I7, meanwhile, tied SOT development directly to accountability for AI impacts: developers ``can be held accountable to what they've done about it because they've just declared that they know.'' Across these accounts, the deliberative work of SOT development generates shared vocabulary and prepares product teams to engage with AI risks; SOT development serves as a ``critical scaffolding'' [C10] for actions intended to mitigate the risks SOT enumerate.

\subsubsection{Interpretation scaffolds action}

SOT scaffold governance actions by reducing the complexity of analysing AI risks. The value of SOT, C10 argued, lies not in the categorisation itself but in the systematic analysis it enables. Across accounts, SOT facilitate the handoff between identifying risks and acting on them [I21, I2, C12, A15, I16, C19, A24, C17]. I16 used the language of `passing' to describe the function of SOT, which enable RAI teams to ``suss out risk types and then pass that off to the [product] team with some helpful advice or strategies that they can use to measure, assess, and determine'' next steps. In these cases, SOT function as translation aids, converting diffuse concerns about AI risks into forms that are tractable for AI product teams, and, in doing so, enabling responsibility for concerns to move from ethics or research teams to product teams. 
This translational function was valued across institutional contexts [A11, I8, I16, I14, I2, C17, C19]: generalised concern about AI risks is, in itself, a barrier to productive action [I4, C17, C12].

A24 extended this scaffolding role in a direction other participants did not, describing SOT as serving a diagnostic function after deployment as well as a proactive one before it: ``If you're building an AI system in any of these areas, this taxonomy helps you pay attention to\ldots\ what can be done to make sure certain types of outcomes are not produced. And on the other flip side\ldots\ [AI] systems are going to make mistakes. And when they do\ldots\ where is the problem? What category of problems are we dealing with?'' A24's account is also distinctive in centring the interests of affected communities rather than the operational needs of product teams: when working with marginalised communities, A24 argued, the right approach is to ``Go talk to the community. Figure out what they need.'' Across our interviews, this orientation was rare; the scaffolding function of SOT was overwhelmingly described in terms of actions for product teams.

\subsubsection{Action without causal analysis may be misdirected}

The form of action SOT scaffold follows from their descriptive approach to enumerating harms. SOT describe AI risks but do not, in general, explain how risks manifest as harms or impacts, or specify the actors whose decisions may exacerbate or mitigate them [I8, I20, C10, C13]. As I8 put it:
\begin{quote}
    There's a missing element of causation in the taxonomy\ldots\ We are treating it like a botany taxonomy as opposed to, `Wait, we're the ones doing this.'\ldots\ Whenever it's discovered or detected and pointed out, of course they\ldots\ go to court and say, `No, we're not responsible.' [I8]
\end{quote}
A botany taxonomy reflects the positivist epistemology of the naturalist classifying plants based on cellular and biological traits, without regard to questions of accountability for their distribution in a particular environment. SOT, treated similarly, render AI harms as objective, passive phenomena to be observed and classified rather than as normative, dynamic outcomes produced by identifiable decisions. As I8 argues, the descriptive posture of existing SOT means that harms are documented without causal attribution, and the organisations producing the underlying systems are positioned to deny responsibility for what SOT record. Other participants similarly highlighted gaps in SOT ``root cause analysis'' [A24]; ``we're only categorising incidents and risks without actually looking at how risks arise'' [I20; cf. C10, I21]. This lack of explicit causal analysis within SOT is, in part, a product of the diffuse range of risk sources unique to the form and function of individual AI systems. I8 notes, when reflecting on conversations about causality with AI product teams during SOT development: ``You'd get to a point in the discussion where they'd say, `Well\ldots\ that's just society, that's just how society is. We can't be held responsible for inequality. We can't be held responsible for racism.''' Where harms can be attributed to the social world in general, rather than to AI development specifically, responsibility for catalogued risks becomes similarly diffuse.
 
Where SOT adopt a descriptive orientation, there are consequences for the kinds of action they can scaffold. Two consequences, in particular, are visible across participants' accounts: action may be misdirected, focusing more on measuring risk proxies than on preventing harm [I20]; and accountability infrastructure that could connect documented harms to responsible actors remains underdeveloped [I7, I8]. The first consequence is visible in how SOT-based measurement work is constructed, while the second can be seen in accountability gaps for addressing SOT-catalogued risks.

A3 described a recurring disconnect between the measurements SOT enable and the impacts SOT enumerate: ``There's a lack of connection made between what is that impact and what is the model or the system's performance?'' Without a causal account that links system features to the impacts they produce, SOT-informed measurement is limited to system properties that are tractable to measure, rather than to the impacts that are the SOT's substantive concern. The action SOT scaffold thus tracks the measurable rather than the consequential. 

The second consequence concerns how actors are held accountable, or made responsible, for addressing risks described in an SOT. I14 described the experience of attempting to challenge a product team's decision-making using SOT-based concerns:
\begin{quote}
    There was one product team that was just driving me absolutely crazy because they didn't really have any justification\ldots\ And there was not really a way to fight that justification because, to them, if they have people buying something that's more justified than, like, a paper [describing an AI risk].
\end{quote}
C12, from civil society rather than within a technology firm, observed: ``You have to have some kind of way to point out what you're talking about [to decision makers].'' These extracts demonstrate an evidentiary and accountability asymmetry. Product teams have tools to articulate how their decisions impact certain forms of consumer behaviour, but RAI teams and civil society lack the tools to articulate how product decisions may increase AI risks in any form that activates internal accountability processes.

Where SOT do inform organisational processes, depending on organisational culture, classification can displace action altogether. I8 provided one example: ``You just kind of point to like, `Oh, we did a datasheet'. Then it's like\ldots\ and what did you do because of that?\ldots\ The answer is, `We didn't actually change it. We just document.''' Datasheets may become evidence that risk has been considered, not evidence that risk has been addressed. I2 named the same risk from a different vantage point, describing the attitude of ``I use this tool and this sort of Bible, in quotes, oh, I'm done\ldots\ It's okay. I got the green light. Push it into market.'' In both accounts, the intended governance work of SOT is displaced by the organisational work of demonstrating governance has occurred. Without causal scaffolding linking categories to decision points and responsible actors, SOT-informed governance may default to dominant logics already operative within the firm.

\subsubsection{Myopic focus on corporate concerns}

The descriptive orientation of SOT can shape which catalogued risks are addressed in practice. Without external requirements to address impacts documented in SOT, decisions about prioritisation may be guided by other incentives, namely a firm's commercial and reputational considerations. I16 described the fundamental tension between near-term ``business needs'' and the ``long time'' it takes to ``understand societal impacts,'' noting that harms often only become ``glaringly apparent after something has already launched.'' Product launches thus act as a contradictory inflection point, marking a transition in the firm's development cycle and the beginning of real-world societal impact. Consequently, at the time initial risk assessments and mitigation strategies are developed, threats to the business and the product timeline are highly legible, while long-term societal impacts remain opaque. RAI practitioners navigate this asymmetry by framing SOT adoption in commercial terms: ``How do we sell this to product teams so they feel like\ldots\ it accelerates their launch'' [I16].

Yet, this commercial framing may influence which SOT-catalogued harms are actually addressed, further ignoring long-term societal impacts. As risks are weighed against business interests, reputational visibility may be a primary proxy for harm. 
As I8 described, the operative question during launch decisions is often simply: ``What would the \textit{New York Times} headline read?'' 
Relying on media coverage and social media monitoring [I7] as the barometer of risk means harms can be systematically deprioritised if they are diffuse, slow-moving, or borne by communities with limited public voice. 
I20 highlighted the geographic bias of this dynamic: ``Europe would focus very much on Europe, US [is] very much concerned about deepfakes and harms associated with that, but then cultural erasure in the Global South kind of gets ignored.''

\subsection{Improving SOT development and adoption}

Across our interviews, participants articulated converging views regarding future development of SOT and their integration into AI governance, highlighting the need for causal analysis linking harm categories to decision points in product development and responsible actors [I8, I20, C13, C12, C10]; external requirements to ensure technology firms are accountable for AI risk mitigation and monitoring [I16, A11, A24, I7]; and coordination across the fragmented landscape of different SOT, moving towards standardised approaches to AI risk classification, identification, and monitoring [C17, A11, I16, I21]. Participants framed technology firms' development of SOT as a pragmatic response to the absence of external requirements and the need for structured approaches to thinking about AI risks and impacts [I16, C17, I8]. This framing reflects the lack of alternatives rather than an endorsement of industry as the appropriate locus for this work.

Participants cautioned, however, that the form that integration takes matters. C17 described the historical trajectory of SOT integration into firm-level AI governance as a flattening movement, from a ``noble minded, ethical, philosophical approach to harm'' through to a ``corporatised risk approach'' integrated into ``normal compliance practices.'' A11 described a sharper version of the same concern, observing that ``bad evals can just serve to increase boldness and reduce potential for liability\ldots\ [and] decrease responsibility for the downstream problems.'' Integration into evaluation and compliance processes, in this account, do not merely flatten but actively reduce accountability, by providing documented evidence that risks have been considered and discharged. The implication is not that further development and adoption of SOT should be resisted, but that steps towards standardisation across SOT should be taken cautiously. Standardisation --- whether through external requirements, internal compliance regimes, or cross-industry coordination --- moves the locus of definitional authority over what counts as an AI risk. Where definitional authority comes to rest shapes which risks remain visible, whose accountability is activated, and what counts as having addressed a classified harm. This highlights the underlying question raised by participants' recommendations: where, within the AI ecosystem, the work of developing and coordinating SOT should sit, whether with regulators [A1], academia [A11], civil society [C10], or distributed across AI governance actors [A3] --- a question we turn to below.

\section{Discussion}


\subsection{Implications for SOT developers and users}

STS scholars have documented how methodological and political choices embedded in classification work are rarely made explicit. This opacity has several consequences for those who inherit the categories \citep{bowker2000SortingThingsOut, star1999LayersSilenceArenasa, suchman2002LocatedAccountabilitiesTechnology}. Phenomena that do not fit are absorbed into residual categories or distorted to fit those that exist. The labour through which the schema was stabilised becomes invisible and difficult to contest, and the positions from which the categories were produced are hidden from those who later use them.

Our findings show this opacity at work in SOT practice, with three consequences revealed in participants' accounts. First, without shared conventions for how SOT are organised, scoped, and described, users cannot reliably compare or combine them. Participants worked around this navigation problem through local mappings that, in aggregate, deepened the problem they were responding to. Second, where the interpretive work informing an SOT is invisible, the categories travel without the caveats that produced them, and SOT designed as reflexive aids can be taken up as performative checklists. Third, where opacity disguises the choices about whose concerns the SOT centres, risks more readily legible to technology firms tend to be reified, while less legible risks, and the communities affected by them, are marginalised. 
We turn now to three design properties that respond to these consequences: interoperability, extensibility, and traceability.

\subsubsection{Interoperability} SOT can be made comparable and combinable through deliberate design, rather than relying solely on after-the-fact translation. Three design moves support this. First, SOT need shared structural conventions through consistent organising primitives: clear distinctions between risks, impacts, and harms; between types of harm and levels at which they occur; and between observed and anticipated effects. Through such primitives, two SOT using compatible conventions can be aligned even if their specific categorical content differs \citep{bowker2000BiodiversityDatadiversity}. Second, explicit metadata regarding evidentiary sources, target audience, intended uses, and definitions of risk and scope should be integrated directly into the SOT rather than accompanying notes. Treating this context as a core component of the artefact makes visible the articulation work of SOT development, exposing the debate, negotiation, and judgement under uncertainty through which categories are stabilised \citep{star1999LayersSilenceArenasa}. This documentation would allow users to assess fit, identify overlap and gaps, and integrate categories across multiple SOT without flattening the vital distinctions that produced them \citep{edwards2011ScienceFrictionData}. Third, the SOT development process itself should produce explicit mappings to adjacent SOT (e.g., noting a category maps to $X$ in another scheme or stating it has no direct equivalent). Publishing SOT with these documented correspondences facilitates interoperability and absorbs the heavy translation labour participants currently report undertaking themselves. 
Practically, as \citet{bagehorn2025ai} describe, this mapping can be supported by adopting stable semantic identifiers for risks, and using frameworks like the Simple Standard for Sharing Ontological Mappings \cite{matentzoglu2022simple}.

\subsubsection{Extensibility} An SOT's categorical structure can leave space for new evidence of risks or harms, technological developments, or contexts to be incorporated.
Three design moves support extensibility \citep{ribes2009LongNowInfrastructure, jackson201411RethinkingRepair}. First, state explicit boundary conditions. Rather than merely documenting what an SOT covers, the artefact should explicitly identify adjacent contexts it \textit{does not} cover and where compatible categorisation is still needed. 
Defining this negative space clearly signals to downstream users where extensions are welcome and roughly what shape they can take. Second, develop risk categories designed to operate across multiple levels of abstraction. When SOT allow categories to be unpacked at finer granularity or aggregated upward without breaking the overall structure, it enables users at different operational altitudes to engage with and extend the artefact in either direction. For example, \citet{shelby2023SociotechnicalHarmsAlgorithmic} organised harms by type and level, providing a modular architecture that \citet{wang2025TaxonomyAlgorithmicHarms} extended to map harms related to disability. Third, include mechanisms for integrating new evidence in the initial publication of the SOT, rather than deferred to future re-development cycles. Conceptually, \citet{weidinger2022TaxonomyRisksPosed} demonstrate this by distinguishing between observed harms and anticipated risks. This built-in distinction creates a designated holding space for new empirical evidence, while the interactive taxonomy developed by \citet{abercrombie2024CollaborativeHumanCentredTaxonomy} offers an example of tooling to support extensibility. 

\subsubsection{Traceability} SOT enumerate harms but, as our third finding documents, rarely connect those harms to the decisions, deployment contexts, and actors implicated in producing them. This disconnect can allow classification to substitute for meaningful action, reducing the potential of SOT to support accountability within AI governance. Two design moves support traceability. First, include causal pathways linking classified risks to the decision points and actors that shape their occurrence. Full causal models require examining sociotechnical specifics of a given AI system and its context of use, and would risk overburdening participants in collaborative SOT development. But even lightweight causal mapping is powerful. By indicating \textit{how} a risk might arise, and \textit{whose} decisions govern that context, the SOT gives users and external stakeholders a concrete basis to challenge what might otherwise be interpreted as diffuse, unavoidable harms. \citet{slattery2026AIRiskRepository}'s Causal Taxonomy, which categorises entity, intent and timing, provides a practical example of this approach. Second, monitoring requirements should be made explicit alongside causal pathways. For each pathway, the SOT should specify what needs to be observable in deployment for a classified risk to be detected. Naming these observability requirements makes the link between classification and intervention explicit, \textit{even when the infrastructure to fulfil them does not yet exist}. A rigorous measurement practice, supported by complementary diagnostic approaches (e.g., \citep{rismani2023, rismani2025measuring, wallach2024EvaluatingGenerativeAI}), will allow SOT developers to surface these observability gaps.

\subsection{Implications for AI governance}


Integration gaps in AI governance reflect more than coordination shortcomings. SOT do important work surfacing potential AI harms across product teams, policymakers, and civil society. But in a regulatory environment with few binding requirements specific to AI harm mitigation, technology firms remain largely able to self-regulate how they respond to surfaced harms \citep{greene2019BetterNicerClearer}. Those outside technology firms have few pathways for holding them to account for risk prioritisation or mitigation actions. Our finding that classification can substitute for action, and can occur without sanction, is unsurprising in this light \citep{costanza-chock2022WhoAuditsAuditors}. The proliferation of SOT compounds the gap. When every firm develops its own internal SOT, harm definition becomes a private rather than a public activity. The consequences of those private definitional choices accrue to publics that have little role in making them.

Where, then, in the AI governance ecosystem should the work of SOT development sit? Participants gestured in different directions, including regulators, academia, civil society, or distributed across actors. Each has structural limits. Industry-led development brings empirical grounding from proximity to deployed systems, but participants described an asymmetric publication regime: researchers in industry labs are rewarded for novel taxonomies and constrained from publishing actual incidents. The result is a steady supply of new SOT alongside little public evidence about whether the harms those SOT enumerate are actually occurring or being addressed. The SOT closest to deployed systems are the least likely to be public, and the most visible SOT are largely categorisation without empiricism. 
Regulator-led development could address concerns about democratic legitimacy --- weighing different harms against each other is, in substance, a political question \citep{bowker2000SortingThingsOut} --- but it risks the flattening of normative commitments into compliance categories that participants observed occurring within firms. Academic and civil society-led development can preserve normative breadth, but lacks the empirical visibility and enforcement leverage that would make SOT operative in deployed contexts. No single locus offers a complete answer.

Standardisation across SOT would, in principle, address many of these challenges by providing a common referent, reducing the strategic space for firm-level selection and partial uptake. But standards emerge through accretive coordination across institutional sites, requiring an accumulated substrate of shared vocabularies, infrastructural commitments, comparable evidentiary records, and consolidated institutional authority \citep{timmermans2010WorldStandardsNot}. The AI governance landscape lacks these. No existing institution combines the empirical visibility, democratic legitimacy, and enforcement leverage that standardisation requires. The technology firms best positioned to lead standard-setting for SOT are also those whose interests standardisation would most readily entrench. Premature standardisation under these conditions risks freezing the definitional asymmetries our findings have identified, on terms that suit incumbents. What is needed first is not a standard but the substrate from which a standard could later emerge.

Assembling this substrate is the AI governance work needed to complement the SOT design recommendations outlined above. A well-designed SOT does little if the conditions for its uptake, comparison, and contestation do not exist; interoperability, extensibility, and accountability must be realised at the level of the ecosystem too. Governance efforts can begin to assemble the infrastructure within which these properties become consequential: accretive infrastructure \citep{timmermans1997StandardizationActionAchieving} through which future standards for AI risk, harm, and impact definition, classification, and monitoring may develop. The most tractable starting point is aggregation: a registry, perhaps hosted by an existing standards institution, where SOT are deposited alongside their developmental documentation and where monitoring and reporting expectations can be attached to catalogued risks. Such a register would do three things. First, it would make interoperability a property of the ecosystem rather than work that each developer replicates. Second, it would offer a place where empirical evidence currently siloed within firms could begin to be assembled, giving extensibility a basis in accumulating evidence rather than in developer foresight alone. Third, it would enable monitoring expectations to be attached to classification work, turning the link between classification and intervention into a potentially enforceable governance relation rather than a developer aspiration. AGORA \citep{arnold2024IntroducingAIGovernance} offers a partial precedent at the regulatory level. This is the broader governance project that SOT scaffold, and on which their value ultimately depends.

\section{Limitations}

Four limitations bear on how our findings should be read. First, our sample mirrors the geographic concentration that our participants themselves critique: participants and authors are based in the US, Europe, UK, and Australia, with no representation from the Global South. The dynamics we describe may look different from positions our team cannot speak from. Second, our findings are based on retrospective accounts of SOT development and use, rather than direct observation of taxonomising practice or its effects on product decisions. Triangulation through ethnographic studies of SOT in use would be helpful. Third, we recruited participants through their prior engagement with SOT, which excludes the perspectives of practitioners who encountered SOT and chose not to use them. Their reasoning would provide vital counter-evidence regarding SOT uptake. Finally, several participants spoke with notable candour and their quotes recur across our themes; we have sought to balance their voices against others where possible, but especially forthright participants likely influenced the analysis.

\section{Conclusion and future work}

We set out to investigate two research questions about the development and use of SOT, specifically their role in AI governance. Through interviews with researchers and practitioners ($n = 25$), we found SOT developers and users to be keenly aware of a core tension: imposing a useful structure on an inherently complex problem space without flattening the nuance required to properly analyse AI risk. While SOT are theoretically positioned to bridge the operationalisation gaps between high-level ethical principles and their implementation in product decisions, we found them to be weakly integrated into actual AI governance processes. Our findings locate this weak integration in features of SOT design and use that simply proliferating more SOT will not, on its own, resolve. The necessary, but reductive, choices through which SOT structure the problem space of AI risks tend to be invisible to downstream taxonomy users, causing frameworks intended as interpretive aids to be misused as exhaustive descriptions. SOT also tend to enumerate harms without linking them to the decision points or specific actors implicated in their occurrence, leaving accountability difficult to assign and facilitating classification to substitute for substantive action.

Realising the governance potential of SOT requires intervention at the artefact and ecosystem levels. For SOT developers, we articulate three desiderata: interoperability, extensibility, and traceability, the last of which demands explicit causal pathways and observability conditions for classified risks. At the ecosystem level, rather than pursuing premature standardisation that risks entrenching incumbent interests, we advocate for assembling the substrate from which future standards might emerge: a shared registry that aggregates SOT alongside their developmental documentation, evidentiary records, and monitoring expectations. Finally, to better understand when classification leads to mitigation, future empirical research must directly observe SOT in use across industry, civil society, and regulatory agencies. 

\section*{Acknowledgements}
We recognise and thank the generosity and trust of our participants, whose contributions enabled the analysis presented. We also extend our thanks to the reviewers for their engagement and constructive feedback. This research project benefited from feedback gathered at the \textit{Evaluating Evaluations: Examining Best Practices for Measuring Broader Impacts of Generative AI} NeurIPS 2024 workshop. The quote included in the title of this paper is from participant I4, who noted that the phrase “\textit{death by a thousand taxonomies}” was first used by Avijit Ghosh and Kevin Klyman during the 2024 NeurIPS workshop.

\bibliography{references}

@inproceedings{hutiri2024NotMyVoice,
  title = {Not {{My Voice}}! {{A Taxonomy}} of {{Ethical}} and {{Safety Harms}} of {{Speech Generators}}},
  booktitle = {The 2024 {{ACM Conference}} on {{Fairness}}, {{Accountability}}, and {{Transparency}}},
  author = {Hutiri, Wiebke and Papakyriakopoulos, Orestis and Xiang, Alice},
  year = 2024,
  month = jun,
  pages = {359--376},
  publisher = {ACM},
  address = {Rio de Janeiro Brazil},
  doi = {10.1145/3630106.3658911}
}

@inproceedings{shelby2023SociotechnicalHarmsAlgorithmic,
  title = {Sociotechnical {{Harms}} of {{Algorithmic Systems}}: {{Scoping}} a {{Taxonomy}} for {{Harm Reduction}}},
  shorttitle = {Sociotechnical {{Harms}} of {{Algorithmic Systems}}},
  booktitle = {Proceedings of the 2023 {{AAAI}}/{{ACM Conference}} on {{AI}}, {{Ethics}}, and {{Society}}},
  author = {Shelby, Renee and Rismani, Shalaleh and Henne, Kathryn and Moon, {\relax Aj}ung and Rostamzadeh, Negar and Nicholas, Paul and {Yilla-Akbari}, N'Mah and Gallegos, Jess and Smart, Andrew and Garcia, Emilio and Virk, Gurleen},
  year = 2023,
  month = aug,
  pages = {723--741},
  publisher = {ACM},
  address = {Montreal QC Canada},
  doi = {10.1145/3600211.3604673},
  isbn = {979-8-4007-0231-0},
  langid = {english}
}

@article{batool2025ai,
  title={AI governance: A systematic literature review},
  author={Batool, Amna and Zowghi, Didar and Bano, Muneera},
  journal={AI and Ethics},
  volume={5},
  number={3},
  pages={3265--3279},
  year={2025},
  publisher={Springer}
}

@inproceedings{rismani2023,
author = {Rismani, Shalaleh and Shelby, Renee and Smart, Andrew and Jatho, Edgar and Kroll, Joshua and Moon, AJung and Rostamzadeh, Negar},
title = {From Plane Crashes to Algorithmic Harm: Applicability of Safety Engineering Frameworks for Responsible ML},
year = {2023},
isbn = {9781450394215},
publisher = {Association for Computing Machinery},
address = {New York, NY, USA},
url = {https://doi.org/10.1145/3544548.3581407},
doi = {10.1145/3544548.3581407},
booktitle = {Proceedings of the 2023 CHI Conference on Human Factors in Computing Systems},
articleno = {2},
numpages = {18},
location = {Hamburg, Germany},
series = {CHI '23}
}

@inproceedings{rismani2025measuring,
  title={Measuring what matters: Connecting AI ethics evaluations to system attributes, hazards, and harms},
  author={Rismani, Shalaleh and Shelby, Renee and Davis, Leah and Rostamzadeh, Negar and Moon, AJung},
  booktitle={Proceedings of the AAAI/ACM Conference on AI, Ethics, and Society},
  volume={8(3)},
  pages={2199--2213},
  year={2025}
}

@inproceedings{weidinger2022TaxonomyRisksPosed,
  title = {Taxonomy of {{Risks}} Posed by {{Language Models}}},
  booktitle = {2022 {{ACM Conference}} on {{Fairness}}, {{Accountability}}, and {{Transparency}}},
  author = {Weidinger, Laura and Uesato, Jonathan and Rauh, Maribeth and Griffin, Conor and Huang, Po-Sen and Mellor, John and Glaese, Amelia and Cheng, Myra and Balle, Borja and Kasirzadeh, Atoosa and Biles, Courtney and Brown, Sasha and Kenton, Zac and Hawkins, Will and Stepleton, Tom and Birhane, Abeba and Hendricks, Lisa Anne and Rimell, Laura and Isaac, William and Haas, Julia and Legassick, Sean and Irving, Geoffrey and Gabriel, Iason},
  year = {2022},
  month = jun,
  pages = {214--229},
  publisher = {ACM},
  address = {Seoul Republic of Korea},
  doi = {10.1145/3531146.3533088},
  urldate = {2022-06-30},
  isbn = {978-1-4503-9352-2},
  langid = {english}
}

@misc{abercrombie2024CollaborativeHumanCentredTaxonomy,
    title = {A {Collaborative}, {Human}-{Centred} {Taxonomy} of {AI}, {Algorithmic}, and {Automation} {Harms}},
    url = {http://arxiv.org/abs/2407.01294},
    doi = {10.48550/arXiv.2407.01294},
    urldate = {2025-11-14},
    publisher = {arXiv},
    author = {Abercrombie, Gavin and Benbouzid, Djalel and Giudici, Paolo and Golpayegani, Delaram and Hernandez, Julio and Noro, Pierre and Pandit, Harshvardhan and Paraschou, Eva and Pownall, Charlie and Prajapati, Jyoti and Sayre, Mark A. and Sengupta, Ushnish and Suriyawongkul, Arthit and Thelot, Ruby and Vei, Sofia and Waltersdorfer, Laura},
    month = nov,
    year = {2024},
    note = {arXiv:2407.01294 [cs]},
}

@techreport{calvino2024SectoralTaxonomyAi,
    title = {A sectoral taxonomy of {AI} intensity},
    url = {https://ideas.repec.org/p/oec/comaaa/30-en.html},
    urldate = {2025-11-14},
    institution = {OECD Publishing},
    author = {Calvino, Flavio and Dernis, Hélène and Samek, Lea and Ughi, Antonio},
    year = {2024},
}

@misc{theofanos2024AIUseTaxonomy,
  author = {Mary Frances Theofanos and Yee-Yin Choong and Theodore Jensen},
  title = {AI Use Taxonomy: A Human-Centered Approach},
  year = {2024},
  month = {2024-03-26 04:03:00},
  publisher = {NIST Trustworthy and Responsible AI, National Institute of Standards and Technology, Gaithersburg, MD},
  url = {https://tsapps.nist.gov/publication/get_pdf.cfm?pub_id=956852},
  doi = {https://doi.org/10.6028/NIST.AI.200-1},
  language = {en},
}

@inproceedings{wallach2024EvaluatingGenerativeAI,
author = {Wallach, Hanna and Desai, Meera and Cooper, A. Feder and Wang, Angelina and Atalla, Chad and Barocas, Solon and Blodgett, Su Lin and Chouldechova, Alexandra and Corvi, Emily and Dow, P. Alex and Garcia-Gathright, Jean and Olteanu, Alexandra and Pangakis, Nicholas and Reed, Stefanie and Sheng, Emily and Vann, Dan and Vaughan, Jennifer Wortman and Vogel, Matthew and Washington, Hannah and Jacobs, Abigail Z.},
title = {Position: evaluating generative AI systems is a social science measurement challenge},
year = {2025},
publisher = {JMLR.org},
booktitle = {Proceedings of the 42nd International Conference on Machine Learning},
articleno = {3318},
numpages = {20},
location = {Vancouver, Canada},
series = {ICML'25}
}

@article{bagehorn2025ai,
  title={AI risk atlas: Taxonomy and tooling for navigating AI risks and resources},
  author={Bagehorn, Frank and Brimijoin, Kristina and Daly, Elizabeth M and He, Jessica and Hind, Michael and Garces-Erice, Luis and Giblin, Christopher and Giurgiu, Ioana and Martino, Jacquelyn and Nair, Rahul and others},
  journal={arXiv preprint arXiv:2503.05780},
  year={2025}
}

@article{matentzoglu2022simple,
  title={A simple standard for sharing ontological mappings (SSSOM)},
  author={Matentzoglu, Nicolas and Balhoff, James P and Bello, Susan M and Bizon, Chris and Brush, Matthew and Callahan, Tiffany J and Chute, Christopher G and Duncan, William D and Evelo, Chris T and Gabriel, Davera and others},
  journal={Database},
  volume={2022},
  pages={baac035},
  year={2022},
  publisher={Oxford University Press UK}
}

@article{slattery2026AIRiskRepository,
  title={The AI risk repository: A meta-review, database, and taxonomy of risks from artificial intelligence},
  author={Slattery, Peter and Saeri, Alexander K and Grundy, Emily AC and Graham, Jess and Noetel, Michael and Uuk, Risto and Dao, James and Pour, Soroush and Casper, Stephen and Thompson, Neil},
  journal={Patterns},
  volume={7},
  issue={5},
  year={2026},
  publisher={Elsevier}
}

@inproceedings{bird2023TypologyRisksGenerative,
    address = {Montreal QC Canada},
    title = {Typology of {Risks} of {Generative} {Text}-to-{Image} {Models}},
    isbn = {9798400702310},
    url = {https://dl.acm.org/doi/10.1145/3600211.3604722},
    doi = {10.1145/3600211.3604722},
    language = {en},
    urldate = {2024-04-16},
    booktitle = {Proceedings of the 2023 {AAAI}/{ACM} {Conference} on {AI}, {Ethics}, and {Society}},
    publisher = {ACM},
    author = {Bird, Charlotte and Ungless, Eddie and Kasirzadeh, Atoosa},
    month = aug,
    year = {2023},
    pages = {396--410},
}

@misc{critch2023TASRATaxonomyAnalysis,
    title = {{TASRA}: {A} {Taxonomy} and {Analysis} of {Societal}-{Scale} {Risks} from {AI}},
    copyright = {Creative Commons Attribution 4.0 International},
    shorttitle = {{TASRA}},
    url = {https://arxiv.org/abs/2306.06924},
    doi = {10.48550/ARXIV.2306.06924},
    urldate = {2025-11-14},
    publisher = {arXiv},
    author = {Critch, Andrew and Russell, Stuart},
    year = {2023},
    note = {Version Number: 2},
}

@techreport{vassilev2024AdversarialMachineLearning,
    address = {Gaithersburg, MD},
    title = {Adversarial machine learning : a taxonomy and terminology of attacks and mitigations},
    shorttitle = {Adversarial machine learning},
    url = {https://nvlpubs.nist.gov/nistpubs/ai/NIST.AI.100-2e2023.pdf},
    number = {NIST 100-2e2023},
    urldate = {2025-11-14},
    institution = {National Institute of Standards and Technology (U.S.)},
    author = {Vassilev, Apostol and Oprea, Alina and Fordyce, Alie and Anderson, Hyrum},
    month = jan,
    year = {2024},
    doi = {10.6028/NIST.AI.100-2e2023},
    pages = {NIST 100--2e2023},
}

@inproceedings{golpayegani2022TaxonomyAIRisks,
    address = {Laguna Hills, CA, USA},
    title = {Towards a {Taxonomy} of {AI} {Risks} in the {Health} {Domain}},
    copyright = {https://doi.org/10.15223/policy-029},
    isbn = {978-1-6654-7184-8},
    url = {https://ieeexplore.ieee.org/document/9951508/},
    doi = {10.1109/TransAI54797.2022.00007},
    urldate = {2025-11-14},
    booktitle = {2022 {Fourth} {International} {Conference} on {Transdisciplinary} {AI} ({TransAI})},
    publisher = {IEEE},
    author = {Golpayegani, Delaram and Hovsha, Joshua and Rossmaier, Leon W. S. and Saniei, Rana and Misic, Jana},
    month = sep,
    year = {2022},
    pages = {1--8},
}

@inproceedings{lee2024DeepfakesPhrenologySurveillance,
    address = {Honolulu HI USA},
    title = {Deepfakes, {Phrenology}, {Surveillance}, and {More}! {A} {Taxonomy} of {AI} {Privacy} {Risks}},
    isbn = {979-8-4007-0330-0},
    url = {https://dl.acm.org/doi/10.1145/3613904.3642116},
    doi = {10.1145/3613904.3642116},
    language = {en},
    urldate = {2025-11-14},
    booktitle = {Proceedings of the {CHI} {Conference} on {Human} {Factors} in {Computing} {Systems}},
    publisher = {ACM},
    author = {Lee, Hao-Ping (Hank) and Yang, Yu-Ju and Von Davier, Thomas Serban and Forlizzi, Jodi and Das, Sauvik},
    month = may,
    year = {2024},
    pages = {1--19},
}

@misc{solaiman2023EvaluatingSocialImpacta,
    title = {Evaluating the {Social} {Impact} of {Generative} {AI} {Systems} in {Systems} and {Society}},
    copyright = {Creative Commons Attribution Share Alike 4.0 International},
    url = {https://arxiv.org/abs/2306.05949},
    doi = {10.48550/ARXIV.2306.05949},
    urldate = {2025-11-14},
    publisher = {arXiv},
    author = {Solaiman, Irene and Talat, Zeerak and Agnew, William and Ahmad, Lama and Baker, Dylan and Blodgett, Su Lin and Chen, Canyu and Daumé, Hal and Dodge, Jesse and Duan, Isabella and Evans, Ellie and Friedrich, Felix and Ghosh, Avijit and Gohar, Usman and Hooker, Sara and Jernite, Yacine and Kalluri, Ria and Lusoli, Alberto and Leidinger, Alina and Lin, Michelle and Lin, Xiuzhu and Luccioni, Sasha and Mickel, Jennifer and Mitchell, Margaret and Newman, Jessica and Ovalle, Anaelia and Png, Marie-Therese and Singh, Shubham and Strait, Andrew and Struppek, Lukas and Subramonian, Arjun},
    year = {2023},
    note = {Version Number: 4},
}

@article{arnold2024IntroducingAIGovernance,
    title = {Introducing the {AI} {Governance} and {Regulatory} {Archive} ({AGORA}): {An} {Analytic} {Infrastructure} for {Navigating} the {Emerging} {AI} {Governance} {Landscape}},
    volume = {7},
    issn = {3065-8365},
    shorttitle = {Introducing the {AI} {Governance} and {Regulatory} {Archive} ({AGORA})},
    url = {https://ojs.aaai.org/index.php/AIES/article/view/31615},
    doi = {10.1609/aies.v7i1.31615},
    urldate = {2025-11-14},
    journal = {Proceedings of the AAAI/ACM Conference on AI, Ethics, and Society},
    author = {Arnold, Zachary and Schiff, Daniel S. and Schiff, Kaylyn Jackson and Love, Brian and Melot, Jennifer and Singh, Neha and Jenkins, Lindsay and Lin, Ashley and Pilz, Konstantin and Enweareazu, Ogadinma and Girard, Tyler},
    month = oct,
    year = {2024},
    pages = {39--48},
}

@misc{cui2024RiskTaxonomyMitigation,
    title = {Risk {Taxonomy}, {Mitigation}, and {Assessment} {Benchmarks} of {Large} {Language} {Model} {Systems}},
    copyright = {Creative Commons Attribution 4.0 International},
    url = {https://arxiv.org/abs/2401.05778},
    doi = {10.48550/ARXIV.2401.05778},
    urldate = {2025-11-14},
    publisher = {arXiv},
    author = {Cui, Tianyu and Wang, Yanling and Fu, Chuanpu and Xiao, Yong and Li, Sijia and Deng, Xinhao and Liu, Yunpeng and Zhang, Qinglin and Qiu, Ziyi and Li, Peiyang and Tan, Zhixing and Xiong, Junwu and Kong, Xinyu and Wen, Zujie and Xu, Ke and Li, Qi},
    year = {2024},
    note = {Version Number: 1},
}

@article{fernandez-macias2022ComprehensiveTaxonomyTasks,
    title = {A {Comprehensive} {Taxonomy} of {Tasks} for {Assessing} the {Impact} of {New} {Technologies} on {Work}},
    volume = {159},
    issn = {0303-8300, 1573-0921},
    url = {https://link.springer.com/10.1007/s11205-021-02768-7},
    doi = {10.1007/s11205-021-02768-7},
    language = {en},
    number = {2},
    urldate = {2025-11-14},
    journal = {Social Indicators Research},
    author = {Fernández-Macías, Enrique and Bisello, Martina},
    month = jan,
    year = {2022},
    pages = {821--841},
}

@book{puscas2023AIInternationalSecurity,
    title = {{AI} and {International} {Security}: {Understanding} the {Risks} and {Paving} the {Path} for {Confidence}-{Building} {Measures}},
    shorttitle = {{AI} and {International} {Security}},
    publisher = {UNIDIR},
    author = {Puscas, Ioana},
    year = {2023},
}

@techreport{TaxonomyHumanRights,
    title = {Taxonomy of Human Rights Risks Connected to Generative AI},
    url = {https://www.ohchr.org/sites/default/files/documents/issues/business/b-tech/taxonomy-GenAI-Human-Rights-Harms.pdf#page=1.37},
    urldate = {2025-11-14},
    institution = {United Nations Human Rights Office of the High Commissioner},
    year = {2023},
    author = {{UN B-Tech team}},
}

@inproceedings{dominguezhernandez2024MappingIndividualSocial,
    address = {Rio de Janeiro Brazil},
    title = {Mapping the individual, social and biospheric impacts of {Foundation} {Models}},
    isbn = {979-8-4007-0450-5},
    url = {https://dl.acm.org/doi/10.1145/3630106.3658939},
    doi = {10.1145/3630106.3658939},
    language = {en},
    urldate = {2025-11-14},
    booktitle = {The 2024 {ACM} {Conference} on {Fairness} {Accountability} and {Transparency}},
    publisher = {ACM},
    author = {Domínguez Hernández, Andrés and Krishna, Shyam and Perini, Antonella Maia and Katell, Michael and Bennett, Sj and Borda, Ann and Hashem, Youmna and Hadjiloizou, Semeli and Mahomed, Sabeehah and Jayadeva, Smera and Aitken, Mhairi and Leslie, David},
    month = jun,
    year = {2024},
    pages = {776--796},
}

@inproceedings{chan2023HarmsIncreasinglyAgentica,
    address = {Chicago IL USA},
    title = {Harms from {Increasingly} {Agentic} {Algorithmic} {Systems}},
    isbn = {979-8-4007-0192-4},
    url = {https://dl.acm.org/doi/10.1145/3593013.3594033},
    doi = {10.1145/3593013.3594033},
    language = {en},
    urldate = {2025-11-14},
    booktitle = {2023 {ACM} {Conference} on {Fairness} {Accountability} and {Transparency}},
    publisher = {ACM},
    author = {Chan, Alan and Salganik, Rebecca and Markelius, Alva and Pang, Chris and Rajkumar, Nitarshan and Krasheninnikov, Dmitrii and Langosco, Lauro and He, Zhonghao and Duan, Yawen and Carroll, Micah and Lin, Michelle and Mayhew, Alex and Collins, Katherine and Molamohammadi, Maryam and Burden, John and Zhao, Wanru and Rismani, Shalaleh and Voudouris, Konstantinos and Bhatt, Umang and Weller, Adrian and Krueger, David and Maharaj, Tegan},
    month = jun,
    year = {2023},
    pages = {651--666},
}

@inproceedings{yampolskiy2016TaxonomyPathwaysDangerous,
    title = {Taxonomy of {Pathways} to {Dangerous} {Artificial} {Intelligence}.},
    url = {https://cdn.aaai.org/ocs/ws/ws0156/12566-57418-1-PB.pdf},
    urldate = {2025-11-14},
    booktitle = {{AAAI} {Workshop}: {AI}, {Ethics}, and {Society}},
    author = {Yampolskiy, Roman V.},
    year = {2016},
    pages = {143--148},
}

@article{slaughter2020AlgorithmsEconomicJustice,
    title = {Algorithms and economic justice: {A} taxonomy of harms and a path forward for the federal trade commission},
    volume = {23},
    shorttitle = {Algorithms and economic justice},
    url = {https://heinonline.org/hol-cgi-bin/get_pdf.cgi?handle=hein.journals/yjolt23&section=12&casa_token=Gd9p6di-n9AAAAAA:IaOp7Jw3Rb-ZINf0HsNOQFOo4zTC3Da_G5rXRIA8IpBPJ09BPi8sNO2vdakiDpziC6cP},
    urldate = {2025-11-14},
    journal = {Yale JL \& Tech.},
    author = {Slaughter, Rebecca Kelly and Kopec, Janice and Batal, Mohamad},
    year = {2020},
    note = {Publisher: HeinOnline},
    pages = {1},
}

@inproceedings{bender2019TypologyEthicalRisks,
    title = {A typology of ethical risks in language technology with an eye towards where transparent documentation can help},
    volume = {1},
    url = {https://faculty.washington.edu/ebender/papers/Bender-Societal-Impact.pdf},
    urldate = {2025-11-14},
    booktitle = {Future of artificial intelligence: language, ethics, technology workshop},
    author = {Bender, Emily M.},
    year = {2019},
}

@book{bowker2000SortingThingsOut,
    title = {Sorting Things Out: Classification and its Consequences},
    isbn = {0-262-52295-0},
    publisher = {MIT Press},
    author = {Bowker, Geoffrey C. and Star, Susan Leigh},
    year = {2000},
}

@incollection{star1989StructureIllStructuredSolutions,
    title = {The {Structure} of {Ill}-{Structured} {Solutions}: {Boundary} {Objects} and {Heterogeneous} {Distributed} {Problem} {Solving}},
    isbn = {978-1-55860-092-8},
    shorttitle = {The {Structure} of {Ill}-{Structured} {Solutions}},
    url = {https://linkinghub.elsevier.com/retrieve/pii/B978155860092850006X},
    language = {en},
    urldate = {2022-08-23},
    booktitle = {Distributed {Artificial} {Intelligence}},
    publisher = {Elsevier},
    author = {Star, Susan Leigh},
    year = {1989},
    doi = {10.1016/B978-1-55860-092-8.50006-X},
    pages = {37--54},
}

@article{rakova2021WhereResponsibleAI,
    title = {Where {Responsible} {AI} meets {Reality}: {Practitioner} {Perspectives} on {Enablers} for {Shifting} {Organizational} {Practices}},
    volume = {5},
    issn = {2573-0142},
    shorttitle = {Where {Responsible} {AI} meets {Reality}},
    url = {https://dl.acm.org/doi/10.1145/3449081},
    doi = {10.1145/3449081},
    language = {en},
    number = {CSCW1},
    urldate = {2022-04-19},
    journal = {Proceedings of the ACM on Human-Computer Interaction},
    author = {Rakova, Bogdana and Yang, Jingying and Cramer, Henriette and Chowdhury, Rumman},
    month = apr,
    year = {2021},
    pages = {1--23},
}

@inproceedings{madaio2020CoDesigningChecklistsUnderstanda,
    address = {New York, NY, USA},
    title = {Co-{Designing} {Checklists} to {Understand} {Organizational} {Challenges} and {Opportunities} around {Fairness} in {AI}},
    url = {https://doi.org/10.1145/3313831.3376445},
    doi = {10.1145/3313831.3376445},
    urldate = {2023-02-26},
    booktitle = {{CHI} '20},
    publisher = {Association for Computing Machinery},
    author = {Madaio, Michael A. and Stark, Luke and Wortman Vaughan, Jennifer and Wallach, Hanna},
    month = apr,
    year = {2020},
    note = {Journal Abbreviation: CHI '20},
    pages = {1--14},
}

@article{star2010ThisNotBoundary,
    title = {This is {Not} a {Boundary} {Object}: {Reflections} on the {Origin} of a {Concept}},
    volume = {35},
    issn = {0162-2439, 1552-8251},
    shorttitle = {This is {Not} a {Boundary} {Object}},
    url = {http://journals.sagepub.com/doi/10.1177/0162243910377624},
    doi = {10.1177/0162243910377624},
    language = {en},
    number = {5},
    urldate = {2022-07-28},
    journal = {Science, Technology, \& Human Values},
    author = {Star, Susan Leigh},
    month = sep,
    year = {2010},
    pages = {601--617},
}

@article{mittelstadt2019AIEthicsToo,
  title={Principles alone cannot guarantee ethical AI},
  author={Mittelstadt, Brent},
  journal={Nature Machine Intelligence},
  volume={1},
  number={11},
  pages={501--507},
  year={2019},
  publisher={Nature Publishing Group UK London}
}

@article{morley2020WhatHowInitial,
    title = {From {What} to {How}: {An} {Initial} {Review} of {Publicly} {Available} {AI} {Ethics} {Tools}, {Methods} and {Research} to {Translate} {Principles} into {Practices}},
    volume = {26},
    issn = {1353-3452, 1471-5546},
    shorttitle = {From {What} to {How}},
    url = {http://link.springer.com/10.1007/s11948-019-00165-5},
    doi = {10.1007/s11948-019-00165-5},
    language = {en},
    number = {4},
    urldate = {2023-02-20},
    journal = {Science and Engineering Ethics},
    author = {Morley, Jessica and Floridi, Luciano and Kinsey, Libby and Elhalal, Anat},
    month = aug,
    year = {2020},
    pages = {2141--2168},
}

@article{jobin2019GlobalLandscapeAI,
    title = {The global landscape of {AI} ethics guidelines},
    volume = {1},
    issn = {2522-5839},
    url = {http://www.nature.com/articles/s42256-019-0088-2},
    doi = {10.1038/s42256-019-0088-2},
    language = {en},
    number = {9},
    urldate = {2022-03-28},
    journal = {Nature Machine Intelligence},
    author = {Jobin, Anna and Ienca, Marcello and Vayena, Effy},
    month = sep,
    year = {2019},
    pages = {389--399},
}

@inproceedings{greene2019BetterNicerClearer,
  title = {Better, {{Nicer}}, {{Clearer}}, {{Fairer}}: {{A Critical Assessment}} of the {{Movement}} for {{Ethical Artificial Intelligence}} and {{Machine Learning}}},
  shorttitle = {Better, {{Nicer}}, {{Clearer}}, {{Fairer}}},
  author = {Greene, Daniel and Hoffmann, Anna Lauren and Stark, Luke},
  year = {2019},
  eprint = {10125/59651},
  eprinttype = {hdl},
  doi = {10.24251/HICSS.2019.258},
  url = {http://hdl.handle.net/10125/59651},
  urldate = {2026-05-20},
  booktitle = {Hawaii {{International Conference}} on {{System Sciences}}},
}

@inproceedings{selbst2019FairnessAbstractionSociotechnical,
    title = {Fairness and abstraction in sociotechnical systems},
    isbn = {978-1-4503-6125-5},
    doi = {10.1145/3287560.3287598},
    booktitle = {Proceedings of the 2019 {Conference} on {Fairness}, {Accountability}, and {Transparency}},
    author = {Selbst, Andrew D. and Boyd, Danah and Friedler, Sorelle A. and Venkatasubramanian, Suresh and Vertesi, Janet},
    year = {2019},
    pages = {59--68},
}

@inproceedings{jacobs2021MeasurementFairness,
    title = {Measurement and {Fairness}},
    isbn = {978-1-4503-8309-7},
    url = {https://doi.org/10.1145/3442188.3445901},
    doi = {10.1145/3442188.3445901},
    booktitle = {Proceedings of the 2021 {ACM} {Conference} on {Fairness}, {Accountability}, and {Transparency}},
    publisher = {Association for Computing Machinery},
    author = {Jacobs, Abigail Z. and Wallach, Hanna},
    year = {2021},
    pages = {375--385},
}

@inproceedings{rajiClosingAIAccountability2020,
    address = {Barcelona Spain},
    title = {Closing the {AI} accountability gap: defining an end-to-end framework for internal algorithmic auditing},
    isbn = {978-1-4503-6936-7},
    shorttitle = {Closing the {AI} accountability gap},
    url = {https://dl.acm.org/doi/10.1145/3351095.3372873},
    doi = {10.1145/3351095.3372873},
    language = {en},
    urldate = {2024-05-14},
    booktitle = {Proceedings of the 2020 {Conference} on {Fairness}, {Accountability}, and {Transparency}},
    publisher = {ACM},
    author = {Raji, Inioluwa Deborah and Smart, Andrew and White, Rebecca N. and Mitchell, Margaret and Gebru, Timnit and Hutchinson, Ben and Smith-Loud, Jamila and Theron, Daniel and Barnes, Parker},
    month = jan,
    year = {2020},
    pages = {33--44},
}

@article{star1989InstitutionalEcologyTranslations,
    title = {Institutional {Ecology}, `{Translations}' and {Boundary} {Objects}: {Amateurs} and {Professionals} in {Berkeley}'s {Museum} of {Vertebrate} {Zoology}, 1907-39},
    volume = {19},
    issn = {0306-3127, 1460-3659},
    shorttitle = {Institutional {Ecology}, `{Translations}' and {Boundary} {Objects}},
    url = {http://journals.sagepub.com/doi/10.1177/030631289019003001},
    doi = {10.1177/030631289019003001},
    language = {en},
    number = {3},
    urldate = {2022-08-23},
    journal = {Social Studies of Science},
    author = {Star, Susan Leigh and Griesemer, James R.},
    month = aug,
    year = {1989},
    pages = {387--420},
}

@article{wang2025TaxonomyAlgorithmicHarms,
    title = {Toward a {Taxonomy} of {Algorithmic} {Harms} for {Disability}: {A} {Systematic} {Review}},
    volume = {8},
    issn = {3065-8365},
    shorttitle = {Toward a {Taxonomy} of {Algorithmic} {Harms} for {Disability}},
    url = {https://ojs.aaai.org/index.php/AIES/article/view/36745},
    doi = {10.1609/aies.v8i3.36745},
    number = {3},
    urldate = {2026-02-05},
    journal = {Proceedings of the AAAI/ACM Conference on AI, Ethics, and Society},
    author = {Wang, Lining and Kameswaran, Vaishnav and Kacorri, Hernisa},
    month = oct,
    year = {2025},
    pages = {2649--2665},
}

@inproceedings{hovy2016SocialImpactNatural,
    title = {The social impact of natural language processing},
    url = {https://iris.unibocconi.it/bitstream/11565/4006472/2/P16-2096.pdf},
    urldate = {2026-02-05},
    booktitle = {The 54th {Annual} {Meeting} of the {Association} for {Computational} {Linguistics} {Proceedings} of the {Conference}, {Vol}. 2 ({Short} {Papers})},
    publisher = {Association for Computational Linguistics},
    author = {Hovy, Dirk and Spruit, Shannon L.},
    year = {2016},
}

@inproceedings{barocas2017ProblemBiasAllocative,
    title = {The problem with bias: {Allocative} versus representational harms in machine learning},
    volume = {1},
    shorttitle = {The problem with bias},
    booktitle = {9th {Annual} {C}onference of the {S}pecial {I}nterest {G}roup for {C}omputing, {I}nformation and {S}ociety},
    publisher = {New York, NY},
    author = {Barocas, Solon and Crawford, Kate and Shapiro, Aaron and Wallach, Hanna},
    year = {2017},
}

@article{timmermans2010WorldStandardsNot,
    title = {A {World} of {Standards} but not a {Standard} {World}: {Toward} a {Sociology} of {Standards} and {Standardization}},
    volume = {36},
    issn = {0360-0572, 1545-2115},
    shorttitle = {A {World} of {Standards} but not a {Standard} {World}},
    url = {https://www.annualreviews.org/doi/10.1146/annurev.soc.012809.102629},
    doi = {10.1146/annurev.soc.012809.102629},
    language = {en},
    number = {1},
    urldate = {2026-02-18},
    journal = {Annual Review of Sociology},
    author = {Timmermans, Stefan and Epstein, Steven},
    month = jun,
    year = {2010},
    pages = {69--89},
}

@article{timmermans1997StandardizationActionAchieving,
  title = {Standardization in {{Action}}: {{Achieving Local Universality}} through {{Medical Protocols}}},
  shorttitle = {Standardization in {{Action}}},
  author = {Timmermans, Stefan and Berg, Marc},
  date = {1997-04},
  year = {1997},
  journal = {Social Studies of Science},
  shortjournal = {Soc Stud Sci},
  volume = {27},
  number = {2},
  pages = {273--305},
  issn = {0306-3127, 1460-3659},
  doi = {10.1177/030631297027002003},
  langid = {english},
}

@inproceedings{klausscheuerman2026TreadingTransparencyTightrope,
  title = {Treading the {{Transparency Tightrope}}: {{A Taxonomy}} of {{Risks}} and {{Benefits}} of {{Foundation Model Data Transparency}} for {{Transparency Advocates}}},
  shorttitle = {Treading the {{Transparency Tightrope}}},
  booktitle = {Proceedings of the 2026 {{CHI Conference}} on {{Human Factors}} in {{Computing Systems}}},
  author = {Klaus Scheuerman, Morgan and Hutiri, Wiebke and Rahmattalabi, Aida and Matthews, Victoria and Xiang, Alice and Andrews, Jerone},
  date = {2026-04-13},
  year = {2026},
  pages = {1--29},
  publisher = {ACM},
  location = {Barcelona Spain},
  doi = {10.1145/3772318.3790364},
  eventtitle = {{{CHI}} 2026: {{CHI Conference}} on {{Human Factors}} in {{Computing Systems}}},
  isbn = {979-8-4007-2278-3},
  langid = {english},
}

@inproceedings{gehrmann2025UnderstandingMitigatingRisks,
  title = {Understanding and {{Mitigating Risks}} of {{Generative AI}} in {{Financial Services}}},
  booktitle = {Proceedings of the 2025 {{ACM Conference}} on {{Fairness}}, {{Accountability}}, and {{Transparency}}},
  author = {Gehrmann, Sebastian and Huang, Claire and Teng, Xian and Yurovski, Sergei and Bhorkar, Arjun and Thomas, Naveen and Doucette, John and Rosenberg, David and Dredze, Mark and Rabinowitz, David},
  date = {2025-06-23},
  year = {2025},
  pages = {2570--2586},
  publisher = {ACM},
  location = {Athens Greece},
  doi = {10.1145/3715275.3732168},
  eventtitle = {{{FAccT}} '25: {{The}} 2025 {{ACM Conference}} on {{Fairness}}, {{Accountability}}, and {{Transparency}}},
  isbn = {979-8-4007-1482-5},
  langid = {english}
}

@inproceedings{li2025CloserLookExisting,
  title = {A {{Closer Look}} at the {{Existing Risks}} of {{Generative AI}}: {{Mapping}} the {{Who}}, {{What}}, and {{How}} of {{Real-World Incidents}}},
  shorttitle = {A {{Closer Look}} at the {{Existing Risks}} of {{Generative AI}}},
  author = {Li, Megan and Bickersteth, Wendy and Tang, Ningjing and Cranor, Lorrie and Hong, Jason and Shen, Hong and Heidari, Hoda},
  date = {2025-10-15},
  year = {2025},
  booktitle = {Proceedings of the AAAI/ACM Conference on AI, Ethics, and Society},
  shortjournal = {AIES},
  volume = {8(2)},
  pages = {1561--1573},
  issn = {3065-8365},
  doi = {10.1609/aies.v8i2.36655}
}

@inproceedings{holstein2019ImprovingFairnessMachine,
    title = {Improving {Fairness} in {Machine} {Learning} {Systems}: {What} {Do} {Industry} {Practitioners} {Need}?},
    isbn = {978-1-4503-5970-2},
    issn = {23318422},
    doi = {10.1145/3290605.3300830},
    booktitle = {Proceedings of the 2019 {CHI} {Conference} on {Human} {Factors} in {Computing} {Systems}},
    author = {Holstein, Kenneth and Vaughan, Jennifer Wortman and Daumé, Hal and Dudík, Miroslav and Wallach, Hanna},
    year = {2019},
    pages = {1--16},
}

@article{madaio2022AssessingFairnessAI,
    title = {Assessing the {Fairness} of {AI} {Systems}: {AI} {Practitioners}' {Processes}, {Challenges}, and {Needs} for {Support}},
    volume = {6},
    issn = {2573-0142},
    shorttitle = {Assessing the {Fairness} of {AI} {Systems}},
    url = {https://dl.acm.org/doi/10.1145/3512899},
    doi = {10.1145/3512899},
    language = {en},
    number = {CSCW1},
    urldate = {2022-04-19},
    journal = {Proceedings of the ACM on Human-Computer Interaction},
    author = {Madaio, Michael A. and Egede, Lisa and Subramonyam, Hariharan and Wortman Vaughan, Jennifer and Wallach, Hanna},
    month = mar,
    year = {2022},
    pages = {1--26},
}

@misc{mitchellModelCard2022,
    title = {Model {Card}},
    url = {https://huggingface.co/bigscience/bloom},
    urldate = {2024-04-03},
    publisher = {Hugging Face},
    author = {Mitchell, Margaret and Pistilli, Giada and Jernite, Yacine and Ozoani, Ezinwanne and Gerchick, Marissa and Rajani, Nazneen and Luccioni, Alexandra Sasha and Solaiman, Irene and Masoud, Maraim and Nikpoor, Somaieh and Ferrandis, Carlos Muñoz and Bekman, Stas and Akiki, Christopher and Contractor, Danish and Lansky, David and McMillan-Major, Angelina and Thrush, Tristan and Ilić, Suzana and Dupont, Gérard and Longpre, Shayne and Dey, Manan and Biderman, Stella and Kiela, Douwe and Baylor, Emi and Le Scao, Teven and Gokaslan, Aaron and Launay, Julien and Muennighoff, Niklas},
    month = jul,
    year = {2022},
}

@article{gebruDatasheetsDatasets2018,
  title={Datasheets for Datasets},
  author={Gebru, Timnit and Morgenstern, Jamie and Vecchione, Briana and Vaughan, Jennifer Wortman and Wallach, Hanna and Iii, Hal Daum{\'e} and Crawford, Kate},
  journal={Communications of the ACM},
  volume={64},
  number={12},
  pages={86--92},
  year={2021},
  publisher={ACM New York, NY, USA}
}

@inproceedings{metcalf2021AlgorithmicImpactAssessments,
    address = {Virtual Event Canada},
    title = {Algorithmic {Impact} {Assessments} and {Accountability}: {The} {Co}-construction of {Impacts}},
    isbn = {978-1-4503-8309-7},
    shorttitle = {Algorithmic {Impact} {Assessments} and {Accountability}},
    url = {https://dl.acm.org/doi/10.1145/3442188.3445935},
    doi = {10.1145/3442188.3445935},
    language = {en},
    urldate = {2022-10-13},
    booktitle = {Proceedings of the 2021 {ACM} {Conference} on {Fairness}, {Accountability}, and {Transparency}},
    publisher = {ACM},
    author = {Metcalf, Jacob and Moss, Emanuel and Watkins, Elizabeth Anne and Singh, Ranjit and Elish, Madeleine Clare},
    month = mar,
    year = {2021},
    pages = {735--746},
}

@article{selbst2021InstitutionalViewAlgorithmic,
    title = {An {Institutional} {View} {Of} {Algorithmic} {Impact} {Assessments}},
    volume = {35},
    issn = {0897-3393},
    number = {1},
    urldate = {2023-01-27},
    journal = {Harv. J. Law Technol.},
    author = {Selbst, Andrew D},
    year = {2021},
}

@inproceedings{costanza-chock2022WhoAuditsAuditors,
    address = {Seoul Republic of Korea},
    title = {Who {Audits} the {Auditors}? {Recommendations} from a {F}ield {S}can of the {A}lgorithmic {A}uditing {E}cosystem},
    isbn = {978-1-4503-9352-2},
    shorttitle = {Who {Audits} the {Auditors}?},
    url = {https://dl.acm.org/doi/10.1145/3531146.3533213},
    doi = {10.1145/3531146.3533213},
    language = {en},
    urldate = {2022-06-30},
    booktitle = {2022 {ACM} {Conference} on {Fairness}, {Accountability}, and {Transparency}},
    publisher = {ACM},
    author = {Costanza-Chock, Sasha and Raji, Inioluwa Deborah and Buolamwini, Joy},
    month = jun,
    year = {2022},
    pages = {1571--1583},
}

@inproceedings{bermanScopingStudyEvaluation2024,
    address = {Honolulu HI USA},
    title = {A {Scoping} {Study} of {Evaluation} {Practices} for {Responsible} {AI} {Tools}: {Steps} {Towards} {Effectiveness} {Evaluations}},
    isbn = {979-8-4007-0330-0},
    shorttitle = {A {Scoping} {Study} of {Evaluation} {Practices} for {Responsible} {AI} {Tools}},
    url = {https://dl.acm.org/doi/10.1145/3613904.3642398},
    doi = {10.1145/3613904.3642398},
    language = {en},
    urldate = {2024-05-19},
    booktitle = {Proceedings of the {CHI} {Conference} on {Human} {Factors} in {Computing} {Systems}},
    publisher = {ACM},
    author = {Berman, Glen and Goyal, Nitesh and Madaio, Michael},
    month = may,
    year = {2024},
    pages = {1--24},
}

@article{veale2021DemystifyingDraftEU,
    title = {Demystifying the {Draft} {EU} {Artificial} {Intelligence} {Act} — {Analysing} the good, the bad, and the unclear elements of the proposed approach},
    volume = {22},
    issn = {2194-4164},
    url = {https://www.degruyterbrill.com/document/doi/10.9785/cri-2021-220402/html},
    doi = {10.9785/cri-2021-220402},
    language = {en},
    number = {4},
    urldate = {2026-02-19},
    journal = {Computer Law Review International},
    author = {Veale, Michael and Zuiderveen Borgesius, Frederik},
    month = aug,
    year = {2021},
    pages = {97--112},
}

@article{kaminski2021AlgorithmicImpactAssessments,
    title = {Algorithmic impact assessments under the {GDPR}: producing multi-layered explanations},
    volume = {11},
    copyright = {http://creativecommons.org/licenses/by-nc-nd/4.0/},
    issn = {2044-3994, 2044-4001},
    shorttitle = {Algorithmic impact assessments under the {GDPR}},
    url = {https://academic.oup.com/idpl/article/11/2/125/6024963},
    doi = {10.1093/idpl/ipaa020},
    language = {en},
    number = {2},
    urldate = {2026-02-19},
    journal = {International Data Privacy Law},
    author = {Kaminski, Margot E and Malgieri, Gianclaudio},
    month = aug,
    year = {2021},
    pages = {125--144},
}

@article{wongSeeingToolkitHow2023,
    title = {Seeing {Like} a {Toolkit}: {How} {Toolkits} {Envision} the {Work} of {AI} {Ethics}},
    volume = {7},
    issn = {2573-0142},
    shorttitle = {Seeing {Like} a {Toolkit}},
    url = {https://dl.acm.org/doi/10.1145/3579621},
    doi = {10.1145/3579621},
    language = {en},
    number = {CSCW1},
    urldate = {2024-03-26},
    journal = {Proceedings of the ACM on Human-Computer Interaction},
    author = {Wong, Richmond Y. and Madaio, Michael A. and Merrill, Nick},
    month = apr,
    year = {2023},
    pages = {1--27},
}

@inproceedings{bietti2020EthicsWashingEthics,
    title = {From {Ethics} {Washing} to {Ethics} {Bashing}: {A} {View} on {Tech} {Ethics} from within {Moral} {Philosophy}},
    isbn = {978-1-4503-6936-7},
    url = {https://doi.org/10.1145/3351095.3372860},
    doi = {10.1145/3351095.3372860},
    booktitle = {Proceedings of the 2020 {Conference} on {Fairness}, {Accountability}, and {Transparency}},
    publisher = {Association for Computing Machinery},
    author = {Bietti, Elettra},
    year = {2020},
    pages = {210--219},
}

@inproceedings{birhane2022PowerPeopleOpportunities,
    address = {New York, NY, USA},
    title = {Power to the {People}? {Opportunities} and {Challenges} for {Participatory} {AI}},
    url = {https://doi.org/10.1145/3551624.3555290},
    doi = {10.1145/3551624.3555290},
    urldate = {2023-04-23},
    booktitle = {{EAAMO} '22},
    publisher = {Association for Computing Machinery},
    author = {Birhane, Abeba and Isaac, William and Prabhakaran, Vinodkumar and Diaz, Mark and Elish, Madeleine Clare and Gabriel, Iason and Mohamed, Shakir},
    month = oct,
    year = {2022},
    note = {Journal Abbreviation: EAAMO '22},
    pages = {1--8},
}

@inproceedings{cooper2022SystematicReviewThematic,
    address = {New Orleans LA USA},
    title = {A {Systematic} {Review} and {Thematic} {Analysis} of {Community}-{Collaborative} {Approaches} to {Computing} {Research}},
    isbn = {978-1-4503-9157-3},
    url = {https://dl.acm.org/doi/10.1145/3491102.3517716},
    doi = {10.1145/3491102.3517716},
    language = {en},
    urldate = {2022-05-02},
    booktitle = {{CHI} {Conference} on {Human} {Factors} in {Computing} {Systems}},
    publisher = {ACM},
    author = {Cooper, Ned and Horne, Tiffanie and Hayes, Gillian R and Heldreth, Courtney and Lahav, Michal and Holbrook, Jess and Wilcox, Lauren},
    month = apr,
    year = {2022},
    pages = {1--18},
}

@book{costanza-chock2020DesignJusticeCommunityled,
    address = {Cambridge, Massachesetts},
    title = {Design justice: community-led practices to build the worlds we need},
    isbn = {978-0-262-35686-2},
    shorttitle = {Design justice},
    language = {eng},
    publisher = {The MIT Press},
    author = {Costanza-Chock, Sasha},
    year = {2020},
    note = {OCLC: 1130310256},
}

@inproceedings{Zhang-dark-side-2025,
    author = {Zhang, Renwen and Li, Han and Meng, Han and Zhan, Jinyuan and Gan, Hongyuan and Lee, Yi-Chieh},
    title = {The Dark Side of AI Companionship: A Taxonomy of Harmful Algorithmic Behaviors in Human-AI Relationships},
    year = {2025},
    isbn = {9798400713941},
    publisher = {Association for Computing Machinery},
    address = {New York, NY, USA},
    url = {https://doi.org/10.1145/3706598.3713429},
    doi = {10.1145/3706598.3713429},
    booktitle = {Proceedings of the 2025 CHI Conference on Human Factors in Computing Systems},
    articleno = {13},
    numpages = {17},
    location = {
    },
    series = {CHI '25}
}

@misc{Alberts-2024-cconversational,
      title={Should agentic conversational AI change how we think about ethics? Characterising an interactional ethics centred on respect}, 
      author={Lize Alberts and Geoff Keeling and Amanda McCroskery},
      year={2024},
      eprint={2401.09082},
      archivePrefix={arXiv},
      primaryClass={cs.CL},
      url={https://arxiv.org/abs/2401.09082}, 
}

@inproceedings{Chandra-lived-experience-2025,
author = {Chandra, Mohit and Naik, Suchismita and Ford, Denae and Okoli, Ebele and De Choudhury, Munmun and Ershadi, Mahsa and Ramos, Gonzalo and Hernandez, Javier and Bhattacharjee, Ananya and Warreth, Shahed and Suh, Jina},
title = {From Lived Experience to Insight: Unpacking the Psychological Risks of Using AI Conversational Agents},
year = {2025},
isbn = {9798400714825},
publisher = {Association for Computing Machinery},
address = {New York, NY, USA},
url = {https://doi.org/10.1145/3715275.3732063},
doi = {10.1145/3715275.3732063},
booktitle = {Proceedings of the 2025 ACM Conference on Fairness, Accountability, and Transparency},
pages = {975–1004},
numpages = {30},
location = {
},
series = {FAccT '25}
}

@inproceedings{Yu-youth-risk-2025,
  title={Youth-Centered GAI Risks (YAIR): A Taxonomy of Generative AI Risks from Empirical Data},
  author={Yaman Yu and Yiren Liu and Jacky Zhang and Yun Huang and Yang Wang},
  booktitle={Symposium On Usable Privacy and Security},
  year={2025},
  url={https://api.semanticscholar.org/CorpusID:280704439}
}

@inproceedings{Lee-privacy-risks-2024,
author = {Lee, Hao-Ping (Hank) and Yang, Yu-Ju and Von Davier, Thomas Serban and Forlizzi, Jodi and Das, Sauvik},
title = {Deepfakes, Phrenology, Surveillance, and More! A Taxonomy of AI Privacy Risks},
year = {2024},
isbn = {9798400703300},
publisher = {Association for Computing Machinery},
address = {New York, NY, USA},
url = {https://doi.org/10.1145/3613904.3642116},
doi = {10.1145/3613904.3642116},
booktitle = {Proceedings of the 2024 CHI Conference on Human Factors in Computing Systems},
articleno = {775},
numpages = {19},
location = {Honolulu, HI, USA},
series = {CHI '24}
}

@misc{Archiwaranguprok-case-informed-harm-2025,
      title={Simulating Psychological Risks in Human-AI Interactions: Real-Case Informed Modeling of AI-Induced Addiction, Anorexia, Depression, Homicide, Psychosis, and Suicide}, 
      author={Chayapatr Archiwaranguprok and Constanze Albrecht and Pattie Maes and Karrie Karahalios and Pat Pataranutaporn},
      year={2025},
      eprint={2511.08880},
      archivePrefix={arXiv},
      primaryClass={cs.HC},
      url={https://arxiv.org/abs/2511.08880}, 
}

@article{Alabed_AI_relationship_2024, title={More than just a chat: A taxonomy of consumers’ relationships with conversational AI agents and their well-being implications}, 
volume={58}, 
rights={https://www.emerald.com/insight/site-policies}, 
ISSN={0309-0566, 0309-0566}, 
url={http://www.emerald.com/ejm/article/58/2/373-409/1226385}, 
DOI={10.1108/EJM-01-2023-0037}, number={2}, journal={European Journal of Marketing}, author={Alabed, Amani and Javornik, Ana and Gregory-Smith, Diana and Casey, Rebecca}, year={2024}, month=feb, pages={373–409}, language={en} }

@misc{Gumusel-privacy-harm-2024,
      title={User Privacy Harms and Risks in Conversational AI: A Proposed Framework}, 
      author={Ece Gumusel and Kyrie Zhixuan Zhou and Madelyn Rose Sanfilippo},
      year={2024},
      eprint={2402.09716},
      archivePrefix={arXiv},
      primaryClass={cs.HC},
      url={https://arxiv.org/abs/2402.09716}, 
}

@article{Zhou-public-health-taxonomy-2025,
author = {Zhou, Jiawei and Chen, Amy Z. and Shah, Darshi and Schwab-Reese, Laura M. and De Choudhury, Munmun},
title = {A Risk Taxonomy and Reflection Tool for Large Language Model Adoption in Public Health},
year = {2025},
issue_date = {November 2025},
publisher = {Association for Computing Machinery},
address = {New York, NY, USA},
volume = {9},
number = {7},
url = {https://doi.org/10.1145/3757544},
doi = {10.1145/3757544},
journal = {Proc. ACM Hum.-Comput. Interact.},
month = oct,
articleno = {CSCW363},
numpages = {32}
}

@misc{zhu-user-discourse-2026,
      title={Understanding Risk and Dependency in AI Chatbot Use from User Discourse}, 
      author={Jianfeng Zhu and Karin G. Coifman and Ruoming Jin},
      year={2026},
      eprint={2602.09339},
      archivePrefix={arXiv},
      primaryClass={cs.CL},
      url={https://arxiv.org/abs/2602.09339}, 
}

@misc{li-user-reported-risk-2025,
      title={Towards Trustworthy AI: Characterizing User-Reported Risks across LLMs "In the Wild"}, 
      author={Lingyao Li and Renkai Ma and Zhaoqian Xue and Junjie Xiong},
      year={2025},
      eprint={2509.08912},
      archivePrefix={arXiv},
      primaryClass={cs.CY},
      url={https://arxiv.org/abs/2509.08912}, 
}

@misc{khoo-minor-bench-2025,
      title={MinorBench: A hand-built benchmark for content-based risks for children}, 
      author={Shaun Khoo and Gabriel Chua and Rachel Shong},
      year={2025},
      eprint={2503.10242},
      archivePrefix={arXiv},
      primaryClass={cs.CL},
      url={https://arxiv.org/abs/2503.10242}, 
}

@article{binns2018FairnessMachineLearning,
  title = {Fairness in {{Machine Learning}}: {{Lessons}} from {{Political Philosophy}}},
  journal = {Proceedings of the 2018 {{Conference}} on {{Fairness}}, {{Accountability}}, and {{Transparency}}},
  author = {Binns, Reuben},
  year = {2018},
  pages = {149--159},
  issn = {23318422},
}

@article{star1999LayersSilenceArenasa,
  title = {Layers of {{Silence}}, {{Arenas}} of {{Voice}}: {{The Ecology}} of {{Visible}} and {{Invisible Work}}},
  shorttitle = {Layers of {{Silence}}, {{Arenas}} of {{Voice}}},
  author = {Star, Susan Leigh and Strauss, Anselm},
  date = {1999-03-01},
  year = {1999},
  journal = {Computer Supported Cooperative Work (CSCW)},
  shortjournal = {Computer Supported Cooperative Work (CSCW)},
  volume = {8},
  number = {1},
  pages = {9--30},
  issn = {1573-7551},
  doi = {10.1023/A:1008651105359},
  langid = {english},
}

@article{bowker2000BiodiversityDatadiversity,
  title = {Biodiversity {{Datadiversity}}},
  author = {Bowker, Geoffrey C.},
  year = {2000},
  journal = {Social Studies of Science},
  shortjournal = {Soc Stud Sci},
  volume = {30},
  number = {5},
  pages = {643--683},
  issn = {0306-3127, 1460-3659},
  doi = {10.1177/030631200030005001},
  langid = {english}
}

@article{edwards2011ScienceFrictionData,
  title = {Science Friction: {{Data}}, Metadata, and Collaboration},
  shorttitle = {Science Friction},
  author = {Edwards, Paul N. and Mayernik, Matthew S. and Batcheller, Archer L. and Bowker, Geoffrey C. and Borgman, Christine L.},
  year = {2011},
  journal = {Social Studies of Science},
  shortjournal = {Soc Stud Sci},
  volume = {41},
  number = {5},
  pages = {667--690},
  issn = {0306-3127, 1460-3659},
  doi = {10.1177/0306312711413314},
  langid = {english},
}

@article{ribes2009LongNowInfrastructure,
  title = {The {{Long Now}} of {{Infrastructure}}: {{Articulating Tensions}} in {{Development}}},
  author = {Ribes, David and Finholt, Thomas A.},
  year = {2009},
  journal = {Journal for the Association of Information Systems (JAIS): Special issue on eInfrastructures},
  shortjournal = {JAIS},
  volume = {10},
  number = {5},
  eprint = {10822/557392},
  eprinttype = {hdl},
  pages = {375--398},
}

@article{jackson201411RethinkingRepair,
  title = {Rethinking Repair},
  author = {Jackson, Steven J.},
  year = {2014},
  journal = {Media Technologies: Essays on Communication, Materiality, and Society},
  pages = {221--39},
  publisher = {MIT Press Cambridge, MA, USA},
}

@article{suchman2002LocatedAccountabilitiesTechnology,
  title = {Located Accountabilities in Technology Production},
  author = {Suchman, Lucy},
  year = {2002},
  journal = {Scandinavian Journal of Information Systems},
  volume = {14},
  number = {2},
  issn = {1901-0990},
}

@inproceedings{narayanan2018TranslationTutorial21,
  title = {Translation Tutorial: 21 Fairness Definitions and Their Politics},
  shorttitle = {Translation Tutorial},
  booktitle = {Proc. Conf. Fairness Accountability Transp., New York, Usa},
  author = {Narayanan, Arvind},
  year = {2018},
  volume = {1170},
  pages = {3},
  url = {https://facctconference.org/static/tutorials/narayanan-21defs18.pdf},
  urldate = {2026-05-18}
}

@misc{knox-harmful-trait-ai-companions-2025,
      title={Harmful Traits of AI Companions}, 
      author={W. Bradley Knox and Katie Bradford and Samanta Varela Castro and Desmond C. Ong and Sean Williams and Jacob Romanow and Carly Nations and Peter Stone and Samuel Baker},
      year={2025},
      eprint={2511.14972},
      archivePrefix={arXiv},
      primaryClass={cs.HC},
      url={https://arxiv.org/abs/2511.14972}, 
}

@ARTICLE{Braun2021Size,
  title        = {One size fits all? {What} counts as quality practice in
                  (reflexive) thematic analysis?},
  author       = {Braun, Virginia and Clarke, Victoria},
  year         = {2021},
  journal = {Qualitative Research in Psychology},
  publisher    = {Informa UK Limited},
  volume       = {18},
  issue        = {3},
  pages        = {328--352},
  date         = {2021-07-03},
  doi          = {10.1080/14780887.2020.1769238},
  issn         = {1478-0887,1478-0895},
  url          = {https://www.tandfonline.com/doi/full/10.1080/14780887.2020.1769238},
  language     = {en}
}

@ARTICLE{Braun2006Using,
  title        = {Using thematic analysis in psychology},
  author       = {Braun, Virginia and Clarke, Victoria},
  year         = {2006},
  journal = {Qualitative Research in Psychology},
  publisher    = {Informa UK Limited},
  volume       = {3},
  issue        = {2},
  pages        = {77--101},
  date         = {2006-01},
  doi          = {10.1191/1478088706qp063oa},
  issn         = {1478-0887,1478-0895},
  url          = {http://www.tandfonline.com/doi/abs/10.1191/1478088706qp063oa},
  language     = {en}
}

@ARTICLE{Byrne2022Worked,
  title        = {A worked example of {Braun} and {Clarke}’s approach to
                  reflexive thematic analysis},
  author       = {Byrne, David},
  year         = {2022},
  journal = {Quality \& Quantity},
  publisher    = {Springer Science and Business Media LLC},
  volume       = {56},
  issue        = {3},
  pages        = {1391--1412},
  date         = {2022-06-01},
  doi          = {10.1007/s11135-021-01182-y},
  issn         = {0033-5177,1573-7845},
  url          = {http://dx.doi.org/10.1007/s11135-021-01182-y},
  language     = {en}
}

\onecolumn

\appendix

\section{Sociotechnical outcomes taxonomies}

\begin{table*}[h]
\centering
\small
\begin{tabularx}{\textwidth}{@{} c X >{\raggedright\arraybackslash}p{3.5cm} @{}}
\toprule
\textbf{Year} & \textbf{Title} & \textbf{Citation} \\ \midrule
2016 & Taxonomy of Pathways to Dangerous Artificial Intelligence & \citep{yampolskiy2016TaxonomyPathwaysDangerous} \\
2019 & A typology of ethical risks in language technology with an eye towards where transparent documentation can help & \citep{bender2019TypologyEthicalRisks} \\
2020 & Algorithms and economic justice: A taxonomy of harms and a path forward for the federal trade commission & \citep{slaughter2020AlgorithmsEconomicJustice} \\
2022 & A comprehensive taxonomy of tasks for assessing the impact of new technologies on work & \citep{fernandez-macias2022ComprehensiveTaxonomyTasks} \\
2022 & Towards a Taxonomy of AI Risks in the Health Domain & \citep{golpayegani2022TaxonomyAIRisks} \\
2022 & Taxonomy of Risks posed by Language Models & \citep{weidinger2022TaxonomyRisksPosed} \\
2023 & Typology of Risks of Generative Text-to-Image Models & \citep{bird2023TypologyRisksGenerative} \\
2023 & Harms from Increasingly Agentic Algorithmic Systems & \citep{chan2023HarmsIncreasinglyAgentica} \\
2023 & TASRA: a taxonomy and analysis of societal-scale risks from AI & \citep{critch2023TASRATaxonomyAnalysis} \\
2023 & AI Risks Taxonomy: Paving the Path for Confidence Building Measures & \citep{puscas2023AIInternationalSecurity} \\
2023 & Sociotechnical Harms of Algorithmic Systems: Scoping a Taxonomy for Harm Reduction & \citep{shelby2023SociotechnicalHarmsAlgorithmic} \\
2023 & Evaluating the Social Impact of Generative AI Systems in Systems and Society & \citep{solaiman2023EvaluatingSocialImpacta} \\
2023 & Taxonomy of Human Rights Risks Connected to Generative AI & \citep{TaxonomyHumanRights} \\
2024 & A Collaborative, Human-Centred Taxonomy of AI, Algorithmic, and Automation Harms & \citep{abercrombie2024CollaborativeHumanCentredTaxonomy} \\
2024 & A sectoral taxonomy of AI intensity & \citep{calvino2024SectoralTaxonomyAi} \\
2024 & Risk taxonomy, mitigation, and assessment benchmarks of large language model systems & \citep{cui2024RiskTaxonomyMitigation} \\
2024 & Mapping the individual, social and biospheric impacts of Foundation Models & \citep{dominguezhernandez2024MappingIndividualSocial} \\
2024 & Not My Voice! A Taxonomy of Ethical and Safety Harms of Speech Generators & \citep{hutiri2024NotMyVoice} \\
2024 & Deepfakes, Phrenology, Surveillance, and More! A Taxonomy of AI Privacy Risks & \citep{lee2024DeepfakesPhrenologySurveillance} \\
2024 & AI Use Taxonomy: A Human-Centered Approach & \citep{theofanos2024AIUseTaxonomy} \\
2024 & Adversarial Machine Learning: A Taxonomy and Terminology of Attacks and Mitigations & \citep{vassilev2024AdversarialMachineLearning} \\
2025 & Introducing the AI Governance and Regulatory Archive (AGORA): An Analytic Infrastructure for Navigating the Emerging AI Governance Landscape & \citep{arnold2024IntroducingAIGovernance} \\
2025 & From Lived Experience to Insight: Unpacking the Psychological Risks of Using AI Conversational Agents & \citep{Chandra-lived-experience-2025} \\
2025 & Understanding and Mitigating Risks of Generative AI in Financial Services & \citep{gehrmann2025UnderstandingMitigatingRisks} \\
2025 & A Closer Look at the Existing Risks of Generative AI: Mapping the Who, What, and How of Real-World Incidents & \citep{li2025CloserLookExisting} \\
2025 & Toward a Taxonomy of Algorithmic Harms for Disability: A Systematic Review & \citep{wang2025TaxonomyAlgorithmicHarms} \\
2025 & The Dark Side of AI Companionship: A Taxonomy of Harmful Algorithmic Behaviors in Human-AI Relationships & \citep{Zhang-dark-side-2025} \\
2026 & Treading the Transparency Tightrope: A Taxonomy of Risks and Benefits of Foundation Model Data Transparency for Transparency Advocates & \citep{klausscheuerman2026TreadingTransparencyTightrope} \\
2026 & AI Risk Repository & \citep{slattery2026AIRiskRepository} \\
\bottomrule
\end{tabularx}%
\caption{A non-exhaustive list of sociotechnical outcomes taxonomies}
\label{tab:published_sots}

\end{table*}

\newpage
\section{Participants}

\begin{table*}[ht]
\centering
\small
\begin{tabularx}{\textwidth}{@{} l >{\raggedright\arraybackslash}X >{\raggedright\arraybackslash}p{3.5cm} l >{\raggedright\arraybackslash}p{4cm} @{}}
\toprule
\textbf{Code} & \textbf{Role} & \textbf{Sector} & \textbf{Region} & \textbf{Inclusion Criteria} \\ 
\midrule
A1 & Research Fellow & Academia & UK & Developed taxonomy \\
I2 & Senior Research Scientist & Industry & EU & Developed taxonomy \\
A3 & PhD Candidate & Academia & North America & Developed taxonomy \\
I4 & Applied Policy Researcher & Industry & North America & Developed taxonomy \\
A5 & PhD Candidate & Academia & North America & Developed taxonomy \\
A6 & PhD Candidate & Academia & North America & Developed taxonomy \\
I7 & Research Scientist & Industry & North America & Developed taxonomy \\
I8 & Senior Research Scientist & Industry & North America & Developed taxonomy \\
A9 & PhD Candidate & Academia & UK & Developed taxonomy \\
C10 & Senior Lecturer & Civil Society/Government & UK & Developed taxonomy \\
A11 & PhD Candidate & Academia & North America & Developed taxonomy \\
C12 & Research Lead & Civil Society/Government & EU & Developed taxonomy \\
C13 & Principal Research Scientist & Civil Society/Government & Australia & Conducts RAI evals \\
I14 & Research Scientist & Industry & North America & Conducts RAI evals \\
A15 & Assistant Professor & Academia & North America & Developed taxonomy \\
I16 & Technical Program Manager & Industry & North America & Conducts RAI evals \\
C17 & Senior Policy Researcher & Civil Society/Government & North America & Conducts RAI evals \\
I18 & Senior Director & Industry & North America & Conducts RAI evals \\
C19 & Senior Technologist & Civil Society/Government & North America & Conducts RAI evals \\
I20 & Research Scientist & Industry & EU & Developed taxonomy \\
I21 & Applied Scientist & Industry & North America & Developed taxonomy \\
C22 & Researcher & Civil Society/Government & North America & Conducts RAI evals \\
C23 & Senior Technologist & Civil Society/Government & North America & Conducts RAI evals \\
A24 & Associate Professor & Academia & North America & Developed taxonomy \\
I25 & Linguist & Industry & North America & Developed taxonomy \\ 
\bottomrule
\end{tabularx}
\caption{Study participants. We recruited participants based on their expertise in responsible AI evaluation, either through developing taxonomies of AI harms or through conducting, managing, or participating in responsible AI evaluations.}
\label{tab:participants_detailed}

\end{table*}

\newpage
\section{Codebook summary}


\begin{longtable}{p{2cm} p{3.2cm} p{6.2cm} p{4.2cm} r}
\label{tab:codebook} \\

\toprule
\textbf{Theme} & \textbf{Sub-group} & \textbf{Description} & \textbf{Example codes} & \textbf{\textit{N}} \\
\midrule
\endhead

\midrule
\multicolumn{5}{p{\textwidth}}{ {\tablename\ \thetable: Codebook: themes, sub-groups, and descriptions. \textit{N} indicates the number of low-level codes within each sub-group. The full codebook is available at \url{https://doi.org/10.5281/zenodo.21830185}. (\textit{Table continues on next page.})}} \\
\endfoot

\bottomrule
\multicolumn{5}{p{\textwidth}}{{\tablename\ \thetable:  Codebook: themes, sub-groups, and descriptions. \textit{N} indicates the number of low-level codes within each sub-group. The full codebook with all codes will be made available as supplementary material.}}
\endlastfoot

\textbf{T1.} Purpose and aims of taxonomies
 & Structuring and clarifying the problem space
 & Taxonomies impose structure on an otherwise overwhelming and ambiguous space of AI consequences. Participants describe taxonomies as providing conceptual clarity, shared vocabulary, and a starting point for navigating the landscape of AI risks and harms.
 & \textit{Conceptual clarity and shared language; Provide a map of harms that lets you locate yourself; Organise information in a systematic way}
 & 20 \\

 & Enabling risk assessment, mitigation, and action
 & Taxonomies enable concrete action by disaggregating abstract concerns into enumerable categories that can be prioritised, evaluated, and mitigated. Participants describe taxonomies as scaffolding for risk assessment, product team deliberation, and intervention design throughout AI development.
 & \textit{Enabling thinking and action on risks; Supporting prioritisation; Formulate mitigation strategies}
 & 20 \\

 & Accountability, transparency, and governance
 & Taxonomies function as governance instruments: enabling documentation, creating accountability opportunities, supporting policy, and making it harder for technology firms to deny knowledge of risks.
 & \textit{To make it harder for tech companies to deny knowledge of risks; Identifying who was harmed and who is responsible; Enabling transparency and judgement of evaluations}
 & 11 \\

 & Community building, participation, and engagement
 & Taxonomies serve community-facing functions: aligning disparate stakeholders, facilitating participation in AI governance, supporting civil society engagement, and reducing barriers for non-technical stakeholders.
 & \textit{Supporting civil society and political engagement; Facilitate participation; Helping build networks in wider community}
 & 8 \\

 & Education, knowledge, and discourse
 & Taxonomies serve educational and knowledge-building functions: helping non-specialists understand risks, highlighting research gaps, expanding discourse, and rebutting claims that harms have been solved.
 & \textit{Help people without a disciplinary background understand risks; Highlighting research gaps; Rebutting people who claim to have solved harms}
 & 8 \\

 & Nature, scope, and characterisation
 & General characterisations of taxonomies, including their procedural value (the development process itself is valuable), their relationship to other artefacts, and aspirations for translating ethical concepts into measurement.
 & \textit{Procedural value of taxonomies; Translate philosophical/ethical issues into concrete measurement and mitigation; Taxonomies are a core ingredient of responsible AI}
 & 16 \\

 & \textit{Other}
 & Residual codes not captured by other sub-groups.
 & \textit{As a tool for practitioners}
 & 1 \\

\midrule

\textbf{T2.} Taxonomy development processes
 &
 & How taxonomies are developed: methods (literature synthesis, stakeholder workshops), team composition (disciplinary backgrounds), scoping decisions (technologies, risk types), and the influence of organisational context, peer review, and audience on design.
 & \textit{Through bringing people together for discussion; Developing taxonomy through synthesis/review of literature; Has to account for peer review pipeline, has to account for what will actually get published}
 & 31 \\

\midrule

\textbf{T3.} Target audiences and stakeholders
 &
 & Who taxonomies are designed for and who uses them: specific audiences (academics, policymakers, developers, evaluators, regulators, civil society, affected communities), audience ambiguity, geographic asymmetries in whose concerns are centred, and divergent mental models across audiences.
 & \textit{No particular defined audience for the paper; People in the Global South trying to situate their experienced harms; Annotators, policy developers, lawyers, ethicists, ethical review managers, all of whom have different mental models}
 & 37 \\

\midrule

\textbf{T4.} Reported uses and adoption
 & Specific use cases and applications
 & Concrete reported uses: categorising incidents, designing evaluations, informing safety analysis, structuring pre-deployment review, contextualising harms in specific domains, translating taxonomies into scorecards or toolkits, and guiding product team deliberation.
 & \textit{Design an evaluation; As part of AI ethics review pre-deployment; Translate into scorecard that model developer or documenter can easily fill out}
 & 42 \\

 & Evidence of adoption and uptake
 & Evidence of taxonomy adoption: citations, policy references, anecdotal reports of internal use within technology firms, uptake by civil society. Evidence is generally thin and indirect.
 & \textit{Anecdotal evidence of use within technology firm by engineers; Referred to in policy discussions; Citations of the paper}
 & 6 \\

 & Adoption factors and barriers
 & Factors affecting adoption: appeal to authority (who authored it), breadth of coverage, integration into tools, pedagogical utility, knowledge requirements, and repetition across taxonomies.
 & \textit{Appeal to authority, i.e.\ who wrote it; Knowledge needed to make use of taxonomies; A lot of copying and repetition in taxonomies that followed early papers}
 & 12 \\

 & Institutional context and how taxonomies are chosen
 & How institutional context shapes taxonomy selection: top-down in government, business-incentive-driven in industry, companies creating their own rather than adopting published ones, distinct `flavours' (academic, enterprise risk management, legal), and the lifecycle from research to soft norms to law.
 & \textit{Companies create their own taxonomies; In industry, it's a combination of business incentives, legal risks, product incentivisation; Research to soft norms to law}
 & 12 \\

 & \textit{Other}
 & Residual codes not captured by other sub-groups.
 & \textit{Taxonomy use case}
 & 1 \\

\midrule

\textbf{T5.} Design features and quality criteria
 & Scope, granularity, and breadth
 & The breadth--depth tension: taxonomies can be too abstract (unintuitive concepts) or too granular (rapid obsolescence). Breadth is valued but must be balanced against usability.
 & \textit{Can be too low level or too high level; If it has too much depth it can become redundant quickly due to technology development; Needs to be granular enough to address my specific area}
 & 11 \\

 & Flexibility, adaptability, and evolution
 & The stability--flexibility tension: taxonomies should be living documents accommodating technological change, but not updated so frequently that users defer action. Extensibility is valued; path dependency is a risk.
 & \textit{A living document that is flexible enough that it can be adapted to new technologies; Creating path dependency by locking in categories of outcomes; Whether or not categories are perfect doesn't matter, just needs to be a flexible starting point}
 & 10 \\

 & Actionability and operationalisability
 & Whether taxonomies enable concrete action. Participants value taxonomies supporting prioritisation, triage, and intervention. Taxonomies reduced to checklists or lacking clarity about intended actions are less useful.
 & \textit{Not allowing you to take action afterwards; If it is used as a checklist, rather than as a tool to inform stakeholder consultation; You need to give people an idea of how to operationalise it}
 & 17 \\

 & Methodological grounding and evidence base
 & How taxonomies are grounded: empirical evidence (incidents, user data, community input), robust methodology, practitioner focus. Taxonomies based on speculation or lacking methodology are critiqued.
 & \textit{Grounded in the problem, based on literature or community; No methodology used to determine categories; Speculation has its place, but foresight of risks needs to come from those affected}
 & 16 \\

 & Category quality and precision
 & Quality of categories: specificity, mutual exclusivity, clarity of definitions, concrete examples, interpretive consistency. Critiques include categories hiding heterogeneous harms, excessive overlap, and lists masquerading as taxonomies.
 & \textit{Categories can hide very different types of harms; Different people can interpret the taxonomy the same way; Categorising harms without defining the harms}
 & 19 \\

 & Conceptual depth and causal reasoning
 & Whether taxonomies enable causal reasoning and trade-off analysis. A key critique is the absence of causal information and the failure to account for technology firms' role in shaping impacts.
 & \textit{Lack of causal information and accounting for role of tech firms in shaping impacts; Failure to consider upstream issues and how they contribute to downstream issues; Enable readers to trace harms from individual level to the societal level}
 & 13 \\

 & Audience fit and accessibility
 & Whether the taxonomy is accessible to its target audience. The tension between technical precision and practitioner-legible language. Academic taxonomies are critiqued as divorced from deployment realities.
 & \textit{Balance accurate terms vs.\ language that practitioners can understand; More academic approaches can be divorced from realities around context of use; Explaining allocative harms in a way that works for CS people}
 & 8 \\

 & \textit{Other}
 & Miscellaneous design considerations: contextual nature of risk, organic adoption as a quality signal, shared language enabling productive disagreement, and domain-specific trade-offs.
 & \textit{Risk is contextual; Shared language allows people to disagree for concrete reasons}
 & 6 \\

\midrule

\textbf{T6.} Framing risks, impacts, and harms
 & Defining and conceptualising risk
 & How participants define risk: likelihood and severity, distinction from hazards and harms, observed vs.\ anticipated risks, domain-specific meanings. Participants note the definitions of risk and harm are confused in the field.
 & \textit{The definitions of risk and harm are confusing in the field; Distinguishing between observed and anticipated risks was important; Risk has a specific meaning in privacy context}
 & 18 \\

 & Defining and conceptualising societal impact
 & How participants conceptualise societal impact: aggregate individual impacts, collective or indirect harms, impacts on social systems or institutions, environmental impacts, and the difficulty of predicting diffuse societal effects.
 & \textit{Collective forms of impact where people are indirectly impacted by a technology; Impact on social systems or institutions; Hard to predict diffuse societal impacts}
 & 14 \\

 & Benefits vs.\ risks framing
 & Whether taxonomies should address benefits alongside risks. Most focus on risks, arguing benefits are already promoted by industry. Risk framing is more legible in governance contexts.
 & \textit{Focus on risk because this is legible in governance spaces; Not interested in taxonomy of benefits---all the consulting companies are doing that; Including benefits that are not just about profit, but about societal positives}
 & 10 \\

 & Risk in technology and deployment context
 & How risk varies by technology and deployment context: general-purpose vs.\ domain-specific systems, modality differences, and lifecycle stages.
 & \textit{Risk and impacts requires a lifecycle approach, integrated into full life cycle of product development; General purpose nature of AI systems makes safe use cases difficult to define}
 & 9 \\

\midrule

\textbf{T7.} Tracking and measuring societal impact
 & Current tracking practices and data sources
 & How impacts are currently tracked: incident databases, user reports, social media, news coverage, trust and safety teams, developer communities, and lawsuits. Tracking is largely reactive and unsystematic.
 & \textit{Through NY Times reporting of AI incidents and reputational risk; Joining developer communities on Discord to understand how models are being used; Currently tracking harms based on lawsuits}
 & 13 \\

 & Desired and ideal tracking approaches
 & What participants wish existed: adverse event reporting pipelines (analogous to pharma), longitudinal studies, independent third-party tracking, and quantification of actual usage patterns.
 & \textit{Would like an adverse event reporting pipeline, but it's a fantasy; Third party similar to a polling agency, tracking impacts using surveys; Would like to understand actual scale of real world impacts}
 & 11 \\

 & Challenges and limitations in tracking
 & Why tracking is difficult: measures focus on model performance not societal impact, diffuse impacts are hard to trace, marginalised communities are underrepresented in incident databases, and tracking may be futile without political will.
 & \textit{Measures focus on system performance rather than societal impact; Pointless unless you have a political movement and infrastructure behind it; In a huge list of AI incidents, very few focused on disabilities---under-reporting}
 & 16 \\

 & Responsibility and infrastructure for tracking
 & Who should be responsible and what infrastructure is needed: distributed responsibility, mandatory reporting modelled on cybersecurity, and the absence of adverse event reporting in technology.
 & \textit{Cybersecurity offers a potential model for rigorous tracking of harms; Pharma companies do adverse event reporting, but no equivalent in tech; Distributed responsibility for tracking societal impact}
 & 9 \\

 & \textit{Other}
 & Residual codes on tracking not captured by other sub-groups.
 & \textit{A university could track long-term impacts of AI on its students}
 & 4 \\

\midrule

\textbf{T8.} Challenges, tensions, and failure points
 & Power dynamics and framing biases
 & Taxonomies embed and reinforce power relations. Dominant framings align with frontier AI company interests. Business incentives override RAI concerns. Launch decisions are driven by PR risk, not harm severity.
 & \textit{RAI taxonomies in tension with business incentives; Industry frame the distribution of risk in a way that suits their interests, focus on technical risks; Decision not to launch a product driven by risk of negative media}
 & 9 \\

 & Adoption and evidence gaps
 & Absence of evidence that taxonomies are effective. Developers report no visibility into use, no formal tracking, frustration with categorisation without action, and scepticism that identifying harm categories is useful in itself.
 & \textit{Have no evidence whether it was useful; Much work on categorisation and less on what to do with taxonomies; Identifying categories of harms might not be that useful in itself}
 & 9 \\

 & Organisational and institutional friction
 & Friction when taxonomies meet organisations: translation gaps across stakeholders with different mental models, limited ethics--product team interaction, challenges publishing from industry, and product teams treating taxonomies as final authority.
 & \textit{Translation gap among stakeholders; Little interaction between ethics team and product teams; Product teams treat them as authoritative, final}
 & 10 \\

 & Misuse and unintended consequences
 & Ways taxonomies can be misused: reducing assessment to documentation exercises, automating harm classification, cherry-picking from proliferating taxonomies, and mistaking abstractions for reality.
 & \textit{Documentation for the sake of documentation; Mistaking the abstract for reality, treating the taxonomy as real; Used to discredit peoples' actual experiences of harms}
 & 7 \\

 & Technical and conceptual difficulties
 & Inherent difficulties: AI is a moving target, diverse perspectives resist consensus, predicting harms is challenging, and general-purpose systems resist fixed categorisation.
 & \textit{Taxonomising a moving target; Everyone has different perspectives and sees the world in different terms; It is easier to taxonomise domains which are being regulated, as lawyers make these things very precise}
 & 13 \\

 & Measurement and monitoring challenges
 & Challenges connecting taxonomy categories to outcomes: no post-deployment tracking, monitoring deprioritised after launch, longitudinal measurement is expensive, and disconnect between model performance metrics and societal impact.
 & \textit{No post-deployment tracking of risks; Monitoring of cumulative impacts is most effective after launch, but is often deprioritsed; Lack of connection between system performance and societal impact}
 & 8 \\

\midrule

\textbf{T9.} Broader AI governance and evaluation landscape
 &
 & Broader observations about AI governance: need for transparency and accountability infrastructure, shift from philosophical to compliance approaches, evaluation rigour, role of tort law and auditing, debate over single vs.\ multiple standard taxonomies, and aspirations for equity-based framings.
 & \textit{Society needs transparency and accountability infrastructure---prerequisite to everything else; Robust evaluation ecosystem would include tort law, making developers responsible for outcomes; Should we have one gold standard taxonomy, or should everyone build their own}
 & 38 \\


\end{longtable}

\twocolumn

\end{document}